\documentclass[12pt,a4paper]{article}
\pdfoutput=1
\usepackage{jheppub}
\usepackage{tikz}
\usepackage{subcaption}
\usepackage{natbib}

\newcommand{\ket}{\rangle}

\begin{document}

\begin{flushright}
	\hfill{OU-TH-1183}
\end{flushright}
\title{\mathversion{bold} Shock waves and delay of hyperfast growth in de Sitter complexity}

 \author{Takanori Anegawa,}
 \author{Norihiro Iizuka}

\affiliation{\it Department of Physics, Osaka University, Toyonaka, Osaka 560-0043, JAPAN}

\emailAdd{takanegawa@gmail.com, iizuka@phys.sci.osaka-u.ac.jp}

\abstract{We study the holographic complexity in de Sitter spacetime, especially how the hyperfast growth of holographic complexity in de Sitter spacetime is affected under a small and early perturbation. The perturbed geometry is de Sitter spacetime with shock waves.~We find that the critical time, at which de Sitter holographic complexity diverges, becomes always greater in the presence of the shock waves, which satisfies the averaged null energy conditions. This means that the hyperfast property of de Sitter complexity is delayed by small perturbations.}

\maketitle

\section{Introduction and summary}
The AdS/CFT correspondence \cite{Maldacena:1997re} is extremely important since it gives the nonperturbative definition of quantum gravity in terms of dual field theory. Through this, we can deepen our understanding of the emergent anti-de Sitter (AdS) spacetime bulk physics. Similarly, we are naturally motivated to establish the holographic framework for de Sitter (dS) spacetime. dS spacetime is a good model of our current universe and describes well the features of the accelerating expansion of the universe. Therefore understanding de Sitter's holographic principle will lead to a better understanding of the structure of quantum theory in our universe.

However, the holographic principle in dS spacetime is not as clearly understood as in AdS. Since AdS and dS spacetime are connected by an analytic connection, it seems that field theory is connected by a similar analytic connection \cite{Strominger:2001pn}. 
In this case, the holographic screen on which the field theory is considered would be the spacelike surface. However the field theory on the space-like surface has an imaginary central charge and has no time, thus its physical meaning is not much clear. Another proposal was made by Susskind \cite{Susskind:2021esx}, where the claim is that the two-dimensional dS spacetime reproduced in the framework of JT gravity is equivalent to the double-scaled Sachdev-Ye-Kitaev (SYK) model \cite{Sachdev:1992fk, Kitaev-talks:2015} in the high temperature. Using the static patch holography, the holographic screen is placed on the stretched horizon which is close to the dS horizon. Then he argues that in dS spacetime, scrambling and the complexity growth must show the hyperfast property. 
Before discussing Susskind's argument, we will briefly comment on two important quantities developed recently: out-of-time ordered correlator (OTOC) and complexity.   

OTOC is one of the good indicators to show whether a theory is chaotic or not and is a correlation function that is not aligned in time order. It is related to the commutator of operators and describes how much the influence of one operator is likely to affect distant operators. In particular, in the holographic setting, this corresponds to considering a black hole geometry with a small energy inserted as a perturbation from the boundary, which induces a shock wave geometry due to the exponential blue shift near the horizon. 
This shock wave destroys the entanglement structure of the thermo-field double state and shows the characteristic behavior of chaos \cite{Shenker:2013pqa}. 
In this way, a black hole exhibits a very large chaotic nature \cite{Sekino:2008he}, which turns out to satisfy the so-called MSS bound \cite{Maldacena:2015waa}. This in turn implies that a field theory that has gravitational dual must show the similar maximally chaotic nature of a black hole. For example, the SYK model reproduces this MSS bound in its low temperature limit \cite{Maldacena:2016hyu, Polchinski:2016xgd}. This is one of the reasons why the SYK model is extremely useful for understanding two-dimensional gravity.

Complexity for a state $|\psi \ket$ on the field theory side is originally the minimum number of gates required to achieve that state $|\psi \ket$ from a reference state. In the holographic setting,  it reflects the geometrical structure of the bulk spacetime, especially it captures the feature of the late time $t$-linear growth of the Einstein-Rosen bridge wormholes \cite{Susskind:2014rva, Susskind:2014moa}.  
There are various proposal for holographic complexity, called complexity = action (CA) \cite{Brown:2015bva, Brown:2015lvg}, CV2.0 \cite{Couch:2016exn}, complexity = volume (CV) \cite{Stanford:2014jda} respectively. These correspond to the full gravitational action of the Wheeler-DeWitt (WdW) patch (CA), 
the spacetime volume of the WdW patch (CV2.0), or the volume of codimension-one extremal surfaces (CV). Qualitative nature of all these proposals at late time match in many situations\footnote{Jackiw-Teitelboim (JT) gravity \cite{Jackiw:1984je,Teitelboim:1983ux} is one of the exception where CV and CA can behave differently \cite{Brown:2018bms,Goto:2018iay,Alishahiha:2018swh,Jian:2020qpp,Cai:2020wpc, An:2018xhv} due to dilaton. This is because the JT vacuum is characterized by both the dilaton and metric and the dilaton gives the difference. See for recent works on JT gravity complexity on dS \cite{Chapman:2021eyy, Anegawa:2023wrk}.}, and recently, it has been pointed out in \cite{Belin:2021bga, Belin:2022xmt} that an infinite class of such observables can exist which might correspond to some arbitrariness of the choice of gates in the field theory complexity definition.

Now let us return Susskind's hyperfast conjecture \cite{Susskind:2021esx}. This claim consists mainly of the following two conjectures. First, if we consider the stretched horizon as a holographic screen, then the behavior of OTOC, {\it i.e.}, scrambling, should be hyperfast in the sense that there is no $\log S$ factor for scrambling time\footnote{If the position at which the operator is inserted is just above the stretched horizon which is very close to the cosmological horizon, the behavior of the OTOC is very different from the one of the chaotic system due to the disappearance of warping factor. Clearly, the behavior of the OTOC depends on where one inserts the operator in dS.}.  Second, the growth rates of complexity must also exhibit a hyperfast behavior. This is naturally understood from the fact that dS spacetime well reflects the properties of the expanding universe. In  \cite{Susskind:2021esx, Jorstad:2022mls}, as we approach a critical time $\tau = \tau_{\infty}$ on the stretched horizon clock, the complexity of dS spacetime behaves as follows, 
\begin{align}
\lim_{\tau \rightarrow \tau_{\infty}}\mathcal{C}_V \to \infty,\ \ \lim_{\tau \rightarrow \tau_{\infty}}\frac{d \mathcal{C}_V}{d\tau} \to \infty.
\end{align}
Thus, the divergence of complexity itself as well as its growth rate can be seen as evidence of the hyperfast property of dS complexity.

In this paper, we will focus on the hyperfast nature of dS complexity, and we study how the small perturbation influences this. In dS bulk, a shock wave is induced by a small and early perturbation. Our goal is to investigate how the property of the hyperfast nature is affected by the shock waves. 
Specifically, we estimate the critical time, at which complexity diverges, based on both the WdW patch and CV proposal. Our WdW patch calculations are for dS$_{3}$, but CV calculations are in dS$_{d+1}$. Regarding dS$_{3}$ WdW patch, since the critical times in both CA and CV2.0 are when the WdW patch reaches the point which has infinite radius (See figure~\ref{fig:dS-Comp1}) \cite{Jorstad:2022mls}, studying the effect of shock waves on WdW patch enable us to evaluate the critical time shift by shock waves in both CA and CV2.0 holographic complexity.  
We also study the critical time shift in CV conjecture, 
and we confirm that both WdW patch and CV calculation give exactly the same answer\footnote{Because qualitative behavior near the critical time ${\tau \rightarrow \tau_{\infty}}$ in all CA, CV2.0, CV conjecture are very similar in general dimensions \cite{Jorstad:2022mls}, and furthermore, in both CA and CV2.0, the critical times are when the WdW patch touches the $r =\infty$, we expect our conclusion should hold in general dimensions.}.

Regarding the WdW patch of dS, the main contribution of the divergence at the critical time ${\tau \rightarrow \tau_{\infty}}$ is the volume divergence at the radius $r$ which becomes large. In both CV2.0 and CA, this means that we only need to examine the depth to which the most future tip of the WdW patch is reached. Interestingly, we find that for dS$_3$ complexity, the critical time $\tau_{\infty}$ delays, {\it i.e.,} $\tau_{\infty}$ becomes always greater after the insertion of the shock waves, which satisfies averaged null energy condition. 
However, the result is opposite in the case of AdS, where critical time\footnote{Although AdS complexity does not exhibit hyperfast growth, nevertheless we can still define the ``critical time'' as when the `tip' of WdW patch reaches $r=0$ of the black hole. See figure  \ref{fig:AdS-Comp1} and \ref{fig:AdS-Comp2}.} becomes faster, {\it i.e.,} $\tau_{\infty}$ becomes lesser after that shock wave perturbation. In the AdS case, this is reflected in the fact that shock waves make the wormholes longer \cite{Shenker:2013pqa}. In some sense, in the dS case, the effects of shock waves are opposite compared with the AdS case, therefore traveling between the north pole to the south pole becomes easier. This can be regarded as  ``time advance'' \cite{Gao:2000ga}.   

The reader might think it a little bit strange that hyperfast is delayed by perturbation in de Sitter spacetime since generically complexity grows under a perturbation. However, in the bulk dS viewpoint, this is very natural. dS spacetime is a spacetime where cosmological constant dominates. However, shock wave makes the universe ``radiation/matter dominance'', instead of cosmological constant dominance. The universe no more exponentially expands if the cosmological constant does not dominate, and  
thus, the nature of hyperfast can be destroyed by inserting a small perturbation in the past.  This is essentially the reason why the critical time   $\tau_{\infty}$ becomes greater in the dS case.  

The organization of this paper is as follows. In \S\ref{sec:shock wave} we review the BTZ black hole with shock waves and also study dS spacetime with shock waves. In this paper, we focus on mainly 2+1 dimensional spacetime. Two methods describing these shock wave geometries are presented and also we check their consistency. In \S\ref{sec:delay} we will look at the response of complexity to perturbations. In particular, we calculate the critical time based on WdW patch calculation in dS$_3$ with shock waves. We also calculate the critical time for the CV. We find that all of them give the same answer.  

Before we go on, we comment on the related literature. The effects of shock waves were studied in relation to quantum information. Especially on the BTZ black hole background, it was studied in \cite{Shenker:2013pqa, Shenker:2013yza, Stanford:2014jda, Roberts:2014isa, Shenker:2014cwa}. On dS background, it was studied in \cite{Aalsma:2020aib, Aalsma:2021kle, Aalsma:2022swk}. Especially in \cite{Aalsma:2020aib}, the effect of a single shock wave in dS spacetime and OTOC was studied, which stimulated us. 

Note added: After we finished this draft, we became aware of \cite{Baiguera:2023tpt}, which presents a related study on the complexity in the presence of shock waves in de-Sitter space. 

\section{BTZ/dS 
spacetime with shock waves
}\label{sec:shock wave}
To understand the effects of shock waves, we will study the geometry of AdS and dS in single shock wave case and double shock waves case. There are two ways that we consider for our analysis.  
\begin{itemize}
\item {\it Method} 1 \\ 
{\it Finding a shock wave geometry on the horizon through the Einstein equation}\\
One way is considering a shock wave on the horizon, which is induced by the delta-function stress tensor. Einstein equation determines the resultant back-reacted metric of the shock wave, and the effect of the shock wave is restricted only on the shock wave, {\it i.e.,} horizon. 
Even though the metric is unchanged by the shock wave except on the horizon, in this case, the two metric coordinates, one above and the other below the shock wave, are connected by a simple translational constant shift on the horizon. 
\item {\it Method} 2 \\ 
{\it Cutting and connecting different solutions}\\   
Another way is preparing the two different geometry, and cutting them at a null surface but that null surface is not necessarily the horizon, and connecting the two geometry at the null surface. 
In this case, the stress tensor is not simply a delta functional. Although it is proportional to a delta function, its magnitude is a function of the radial coordinate. 
The difference of coordinates at the connecting null surface is not as simple as above {\it Method} 1. However, in certain limits, one can see the consistency with the {\it Method} 1.
\end{itemize}
Hence we will analyze shock waves following these two methods for the BTZ case and dS case respectively. We will also check the consistency between {\it Method} 1 and 2 as well. 

\subsection{BTZ spacetime}

We first consider the case of BTZ with shock waves. 

The BTZ black hole metric is 
\begin{align}
&ds^2=-f(r)dt^2+\frac{dr^2}{f(r)}+r^2 d\phi^2,\\
&f(r)=\frac{r^2-R^2}{\ell^2}, 
\end{align}
where $f(r)$ is blackening factor and $\ell$ is AdS-radius. $R$ is the horizon radius which is related to the  BTZ black hole mass  $M$ as 
\begin{align}
R^2=8G_N M \ell^2 \,.
\end{align} 
Let us rewrite this in Kruskal coordinates defined by
\begin{align}
\label{KruskalRBTZ}
U_R=-e^{-\frac{R}{\ell^2}(t_R-r^{\ast})},\ V_R=e^{\frac{R}{\ell^2}(t_R+r^{\ast})},
\end{align}
where $r^{\ast}(r)$ is a tortoise coordinate, 
\begin{align}
\label{tortoiseAdS}
r^{\ast}(r)=\int \frac{dr}{f(r)}=\frac{\ell^2}{2R}\log \left| \frac{r-R}{r+R}\right| \,.
\end{align}
The subscript $R$ in $t_R$ indicates that this definition only covers the right Rindler patch. At the black hole horizon $r \to R$, $r^{\ast}(r) \to - \infty$. 
To describe the black hole patch, the left patch, and the white hole patch, 
we need to perform analytic continuation for $t$ in eq.~\eqref{KruskalRBTZ} as 
\begin{align}
t \to t+ \frac{i \beta}{4}\,, \quad t\to t+ \frac{i \beta}{2} \,, 
\quad t\to t+ \frac{3 i \beta}{4} \,, 
\quad \mbox{where} \,\,\, \beta = \frac{2 \pi \ell^2}{R}
\end{align} 
respectively. In this coordinate, the BTZ metric is
\begin{align}
\label{BTZmetric}
ds^2=\frac{-4\ell^2dUdV}{(1+UV)^2}+R^2 \left( \frac{1-UV}{1+UV} \right)^2 d\phi^2 \,,
\end{align}
and the black hole horizon is $U=0$ or $V=0$. 

\begin{figure}[h]
\centering
\includegraphics[keepaspectratio,scale=0.5]{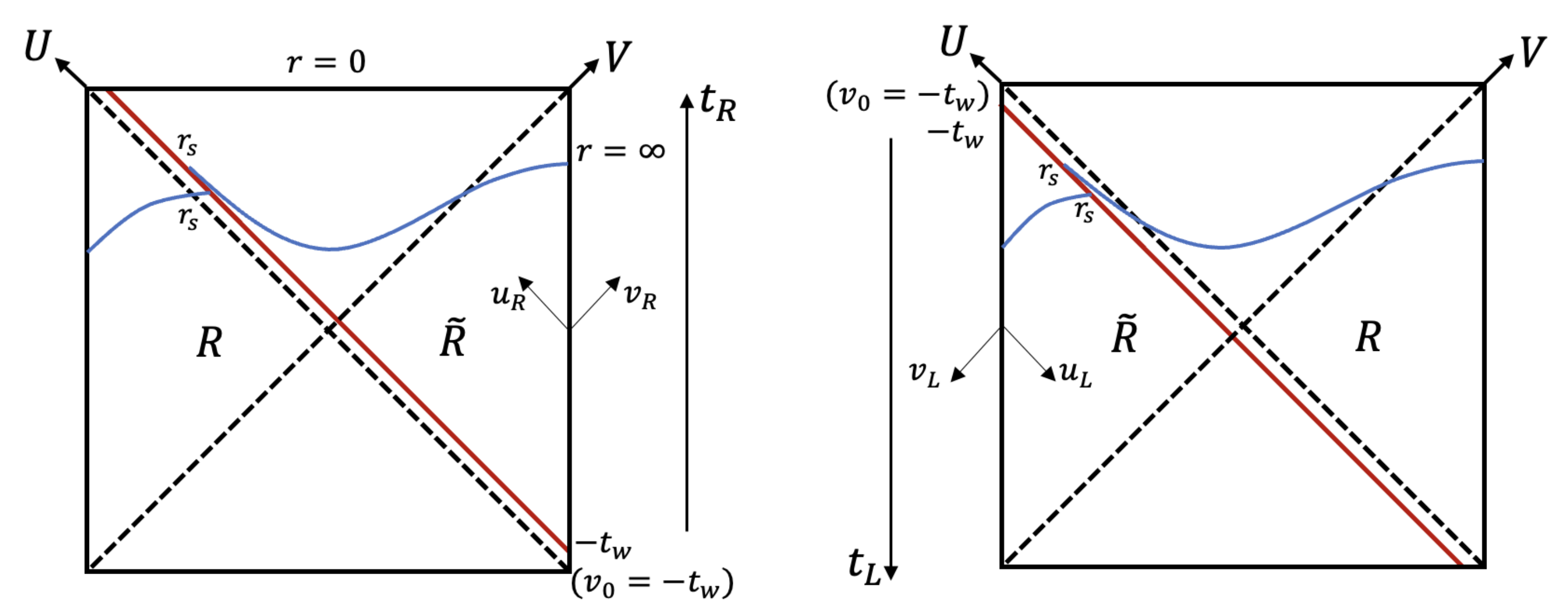}
\caption{Penrose diagram of single shock wave on BTZ geometry. Here we consider the case where a single shock wave is inserted at $t = -t_w$ on the right or left boundary. The horizons are $R$ and $\tilde{R}$ before and after the shock wave passes through. Here before and after are for the right (left) boundary time, $t_R = - t_w$ ($t_L = - t_w$) for the left (right) figure. On this geometry, a geodesic jump at the shock wave. 
For example, a blue geodesic starting from the left boundary intersects a shock wave at $r=r_s$, but it emerges from the same $r=r_s$ but different $U$, because $r$ is continuously connected to the other geometry but $U$ is not.}
\label{fig:AdS-coordinate}
\end{figure}

One can define the null coordinates $(u,v)$ as follows.
\begin{align}
\label{uvRforAdS}
u_R=t-r^{\ast},\ v_R=t+r^{\ast} \,.
\end{align}
Again, $R$ stands for the right patch. On the left patch, we can define
\begin{align}
u_L=t-r^{\ast},\ v_L=t+r^{\ast} \,,
\end{align}
where on the left, $t$ goes backward.  
Unlike $(U, V)$ coordinates, these $(u_R, v_R)$ null coordinates are not defined smoothly beyond the horizon and thus cover only the right patch in Figure \ref{fig:AdS-coordinate}.  
Note that on the right boundary, $r^{\ast} = 0 $ and $u_R = v_R = t$ ranges from $-\infty$ to $\infty$. 

The ingoing Eddington-Finkelstein coordinates are   
\begin{align}
ds^2&=-f(r)dv_R^2 +2dv_Rdr+r^2 d\phi^2  
\end{align}
where $(v_R, r)$ coordinates cover the right and black hole patch. 
See figure \ref{fig:AdS-coordinate}. \\

\subsubsection{Single shock wave}
\begin{figure}[h]
\centering
\includegraphics[keepaspectratio,scale=0.25]{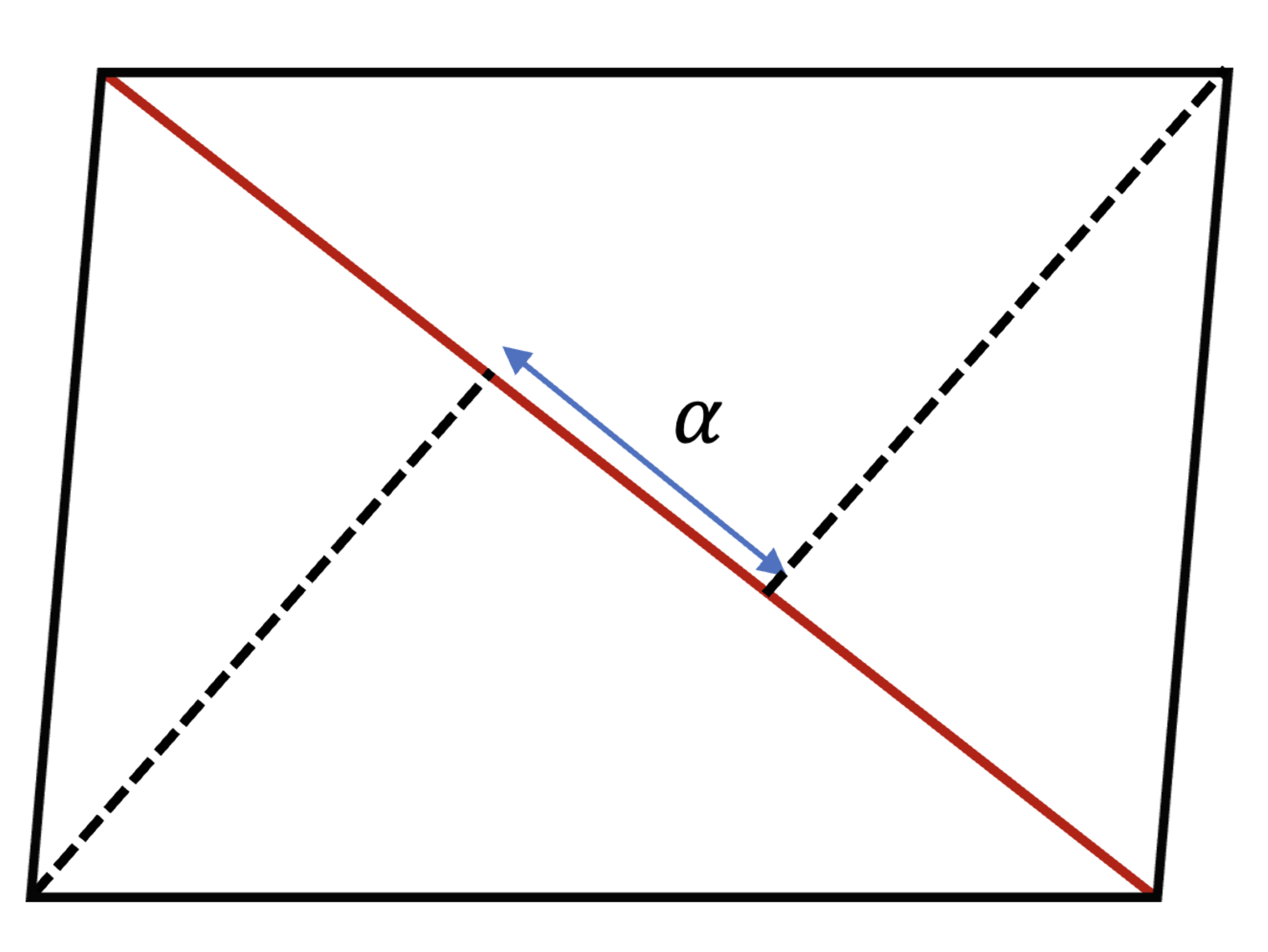}
\caption{Penrose diagram of BTZ geometry with single shock wave which is inserted on the horizon. The red line represents a shock wave. In this case, a time delay of $\alpha > 0$ is generated. This can be interpreted as a simple constant shift only when the shock wave is as close as possible to the horizon.}
\label{fig:AdSshockPen}
\end{figure}

We would now like to study the BTZ black hole metric with a single shock wave. As we mentioned, we study the shock wave geometry based on two ways;  \\
{\it Method} 1\\
Suppose a shock wave is launched along $V=0$. Then we want to find the solution of the Einstein Equation with the following stress tensor, 
\begin{align}
\label{STensor1}
T_{VV}=\frac{\alpha}{4 \pi G_N}\delta(V) \,,
\end{align}
where $\alpha > 0$ due to the averaged null energy condition (ANEC).  
The resultant metric is as follows
\begin{align}
ds^2=\frac{-4\ell^2dUdV}{(1+UV)^2}+4\ell^2 \alpha \delta(V)dV^2+R^2 \left( \frac{1-UV}{1+UV} \right)^2 d\phi^2.
\label{shockgeo12}
\end{align}
Clearly, this metric is the same as eq.~\eqref{BTZmetric} except $V=0$. 
It is convenient to rewrite this metric as follows.
\begin{align}
\label{shockgeo1}
ds^2=\frac{-4\ell^2dUdV}{(1+(U+\alpha \theta(V))V)^2}+R^2 \left( \frac{1-(U+\alpha \theta(V))V}{1+(U+\alpha \theta(V))V} \right)^2 d\phi^2
\end{align}
Setting $U \to U - \alpha \theta(V)$,  eq.~\eqref{shockgeo1} becomes \eqref{shockgeo12}, therefore they are equivalent. 
However written the metric as eq.~\eqref{shockgeo1}, the spacetime structure becomes clearer.  
The spacetime is divided into two parts by the shock wave which is on the horizon $V=0$. The metric is unmodified by the shock wave at $V>0$ and $V<0$, namely if $V=0$ is avoided, there is no effect of the shock wave.
However, at $V=0$, $U$ coordinate is shifted by the amount of $\alpha > 0$. 
See Figure \ref{fig:AdSshockPen} for the Penrose diagram. 
These two black holes have the same mass, and 
this corresponds to inducing a time delay on the horizon. As a result, the distance between the left boundary to the right boundary becomes longer by the shock wave \cite{Shenker:2013pqa} as butterfly effects. 

$\hspace{-7.5mm}${\it Method} 2\\
The second way is preparing the black hole geometry with different masses. In this case, the shock wave is not necessary on the horizon. For example, we can consider the case that a shock wave is released at the time $-t_w$ from the right/left boundary. More concretely, we consider the situation where the mass of the BTZ black hole is modified from $M$ to $M+E$ by the insertion of a shock wave. This has a natural interpretation that the mass has increased due to the insertion of a shock wave. Each black hole is described by $(U,V)$ and $(\tilde{U},\tilde{V})$ coordinates respectively. See figure \ref{fig:AdS-coordinate}.

In this case, the shock wave trajectory can be described in $UV$ coordinates as 
\begin{align}
\label{sshockgeoV}
V=\pm e^{-\frac{Rt_w}{\ell^2}},\ \tilde{V}=\pm e^{-\frac{\tilde{R}t_w}{\ell^2}}
\end{align}
where we take the positive sign for the situation in the left figure in figure \ref{fig:AdS-coordinate}, and the negative sign for the one in the right figure. In the $uv$ coordinate system, it is given as a constant surface where 
\begin{align}
v_{R(L)}=v_0 \equiv - t_w \,.
\end{align}
The $(U, V)$ and $(\tilde{U},\tilde{V})$ coordinates are connected at this shock wave trajectory, and $r$ must be continuous there such that the metric is $C^0$. 
The smoothness of $r$ at the shock wave trajectory requires 
\begin{align}
\label{bdrycon}
R \frac{1-UV}{1+UV}=\tilde{R}\frac{1-\tilde{U}\tilde{V}}{1+\tilde{U}\tilde{V}}
\end{align}
where $V$ and $\tilde{V}$ are given by eq.~\eqref{sshockgeoV} and 
\begin{align}
\label{RtRrela}
R^2 =8G_N M \ell^2 \,, \quad \tilde{R}^2=8G_N \left( M  + E \right) \ell^2.
\end{align}

Thus, the metric is $C^0$ and it can also be written in the Vaidya metric as
\begin{align}
\label{Vaidya1}
ds^2 &=-\frac{1}{\ell^2}(r^2-R^2\theta(v_0-v)-\tilde{R}^2\theta(v-v_0))dv^2+2dv dr+r^2 d\phi^2 \\
&=-\frac{1}{\ell^2}(r^2-R^2 -8 G_N E  \ell^2 \theta(v-v_0))dv^2+2dv dr+r^2 d\phi^2
\end{align}
Here $v$ means $v_R$ ($v= v_L$) in the left (right) figure \ref{fig:AdS-coordinate}. 
This is the solution of the Einstein equation under the following stress tensor.
\begin{align}
\label{STensor2}
T_{vv}=\frac{E}{2 \pi r}\delta(v-v_0) \,.
\end{align}
The ANEC requires $E > 0$. 
This shock wave has some strange properties since it is induced by the stress tensor whose magnitude depends on $r$. However, as we will see below, this is consistent with {\it Method} 1 in certain limits. \\
{\it \underline{Consistency check}}\\
We now check the consistency that the metric obtained with {\it Method} 2 matches the one obtained with  {\it Method} 1, {\it i.e.,} the shock wave geometry eq.~\eqref{shockgeo1} in a certain limit. 

As described earlier, in {\it Method} 2, the shock wave propagates not necessarily on the horizon. To see the consistency with the metric obtained in  {\it Method} 1, 
we need to take the limit large 
$t_w$
so that the shock wave approaches the horizon. 
Furthermore,  in  {\it Method} 1, the geometry is unmodified except for the horizon. This implies that $E$ must be small. 

Thus we are led to the following double scaling limit, 
\begin{align}
\label{BTZDS}
t_w \to \infty \,, \quad \frac{E}{M} \to 0 \,, \quad \mbox{where}  \quad \frac{E}{M} e^{\frac{R}{\ell^2}t_w} = \mbox{fixed} \,.
\end{align}
Under this  double scaling limit, using eq.~\eqref{sshockgeoV}, \eqref{bdrycon} and \eqref{RtRrela}, we can obtain the relationship between $U$ and $\tilde{U}$, which turns out  
\begin{align}
\label{UtildeUrela}
\tilde{U} = U \pm \alpha, \quad  \alpha =\frac{E}{4M}e^{\frac{R}{\ell^2}t_w} > 0 \,.
\end{align}
Here $+ (-)$ is for the left (right) figure \ref{fig:AdS-coordinate}. 
This corresponds to the extreme case where the shock wave is localized on the horizon and $\tilde{R} \to R$. 
This is simply a black hole geometry with the same mass connected by constant shift translation obtained in {\it Method} 1.

Note that in the case where the shock wave is inserted on the left boundary at $t_L = -t_w$, the region where the tilde coordinates are defined is the exact opposite of the case where the shock wave is inserted on the right boundary at $t_R = -t_w$, as shown in figure \ref{fig:AdS-coordinate}. 
Then $U$ and $\tilde{U}$ are switched and 
eq.~\eqref{UtildeUrela} is exactly what we obtained in the metric eq.~\eqref{shockgeo1}.
In this way, the shock wave geometry inserted at time $-t_w$ from the right/left boundary gives essentially the same shift between different coordinates $U$ and $\tilde{U}$ under the double scaling limit eq.~\eqref{BTZDS}.
 
The stress tensor eq.~\eqref{STensor2} is also consistent with eq.~\eqref{STensor1}. To see this, we have 
\begin{align}
T_{VV}&=\left( \frac{dv}{dV}\right)^2 T_{vv}(v)
\,, \quad \frac{dV}{dv}=\frac{R}{\ell^2}e^{\frac{R}{\ell^2}v} 
\end{align}
and in the double scaling limit $v = v_0 = - t_w \to -\infty$, $r \to R$ and 
\begin{align}
 \delta(v-v_0) = \left( \frac{dv}{dV}\right)^{-1}  \delta(V - e^{-\frac{R}{\ell^2} t_w} )\to \left( \frac{dv}{dV}\right)^{-1}  \delta(V )
\end{align}
thus, starting from eq.~\eqref{STensor2}, 
\begin{align}
T_{VV} &= \left( \frac{dv}{dV}\right)^2 \frac{E}{2 \pi r} \delta (v-v_0) \nonumber \\
&\to    \left( \frac{dv}{dV}\right) \frac{E}{2 \pi R} \delta (V)  =\frac{E}{16G_N \pi M}e^{\frac{R}{\ell^2}t_w}\delta(V) 
=\frac{\alpha}{4\pi G_N}\delta(V) \,.
\end{align}
we obtain  eq.~\eqref{STensor1}. 
%
\subsubsection{Double shock waves}
\label{sec:2.2.1}
Next, we consider the case where two shock waves are inserted. We will describe the metric based on the two previous methods. \\
{\it Method} 1\\
First, we consider the case where two shock waves pass over two horizons. {\it i.e.}, $V=0$ and $U=0$, where 
the following stress tensors are inserted,  
\begin{align}
T_{VV}=\frac{\alpha}{4\pi G_N }\delta(V) \,, \quad T_{UU}=\frac{\alpha}{4\pi G_N } \delta(U) \,.
\end{align}
Just like the single shock wave case, we can find the following solution
\begin{align}
\label{twodeltashock}
ds^2=\frac{-4\ell^2dUdV}{(1+UV)^2}+4\ell^2 \alpha \delta(U)dU^2+4\ell^2 \alpha \delta(V)dV^2+R^2 \left( \frac{1-UV}{1+UV} \right)^2 d\phi^2.
\end{align}
However, this is a solution only in the leading order of $\alpha$, neglecting $O(\alpha^2)$. On the other hand, the single shock wave solution eq.~\eqref{shockgeo12} is an exact solution. Therefore we assume $\alpha$ is small enough to justify this perturbative solution in $\alpha$. 

At constant  $U (\neq 0)$ surface, (or constant $V (\neq 0)$ surface), this geometry is a solution with a single shock wave at $V=0$  (or $U=  0$). Therefore, if we leave aside the points where the shock waves intersect, we will see a shift of coordinates on each shock wave, as seen in the single shock wave solution. \\
{\it Method} 2\\
Next, let us consider the method connecting two black hole geometries with different masses. The same continuity condition for $r$ as a single shock wave case must be imposed across shock waves. 

However, a nontrivial additional condition is required at the point where shock waves intersect each other, which is Dray-’t$\,$Hooft-Redmount (DTR) condition \cite{Dray:1985yt, Redmount:1985pr, Poisson:1990eh}. DTR condition is the consequence of the $C^0$ property of metric at the intersection point of delta-functional shock waves. Suppose dividing the spacetime into 4 regions at the intersection of shock waves, a(bove), b(elow), r(ight), and l(eft) of the intersection in the Penrose diagram, then the blackening factor for $f_{a}$, $f_{b}$, $f_{r}$, $f_{l}$ must satisfy 
\begin{align}
\label{DTRcondition}
f_a(r_c)f_b(r_c)=f_l(r_c)f_r(r_c)  
\end{align} 
at the shock wave intersection radius $r=r_c$. 

\begin{figure}[h]
\centering
\includegraphics[keepaspectratio,scale=0.4]{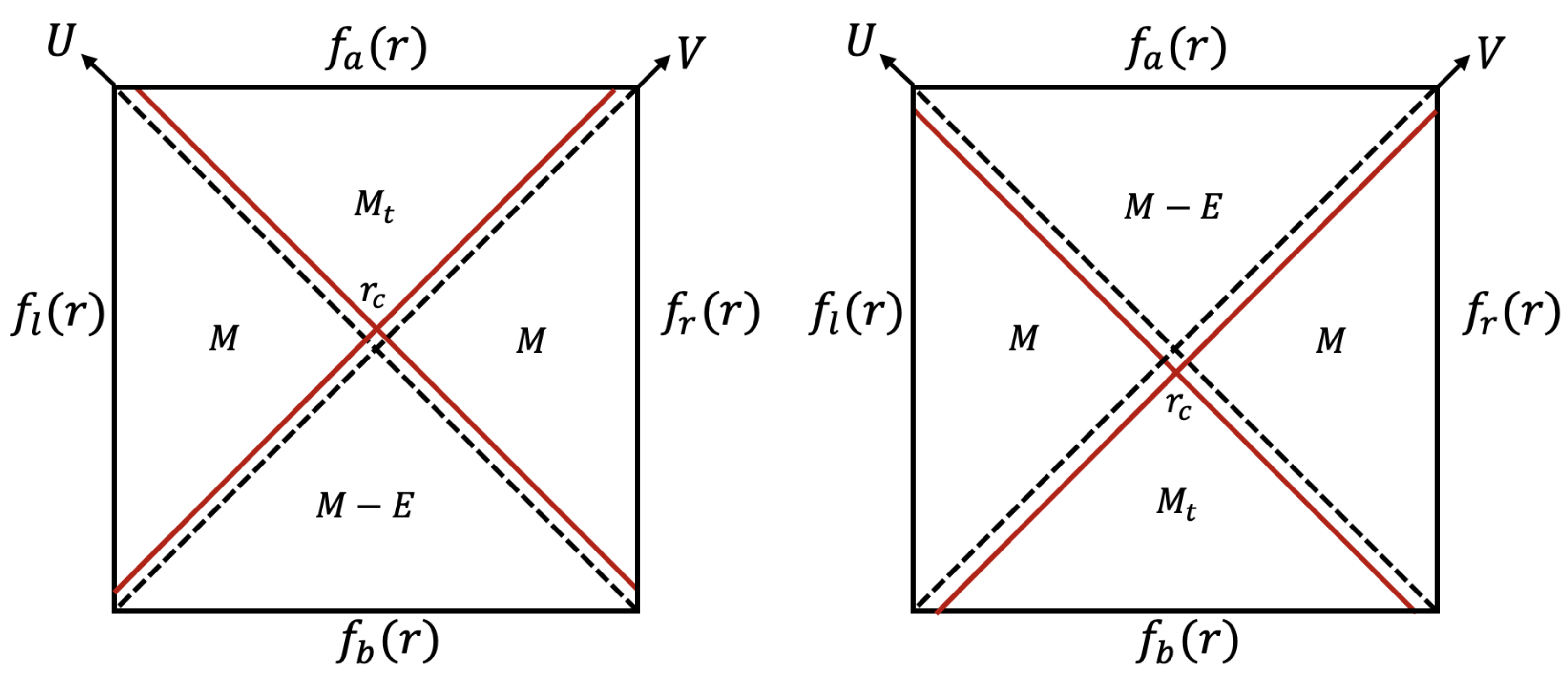}
\caption{BTZ space-time divided by two shock waves (red solid lines). For simplicity, we consider the insertion of double shock waves in such a way that the Penrose diagram is left-right symmetrical. There is a choice to be made as to which boundary to insert the shock wave from $t=-t_w$. In the left figure, one shock wave is inserted at the right boundary time $t_R = -t_w$. 
Another shock wave reaches the left boundary at $t_L = + t_w$. 
In the right figure, the situation is the other way around; one shock wave reaches the right boundary at $t_R = + t_w$, and one shock wave is inserted at $t_L = - t_w$ on the left boundary. 
Shock waves divide the spacetime into four distinguished region as $a$(bove), $b$(elow), $l$(eft), $r$(ight), and their masses are $M_t$, $M-E$, $M$ and $M$ respectively in the left figure. Two shock waves intersect at $r=r_c$.}
\label{fig:AdSdshock}
\end{figure}

Let us consider this condition more concretely. Suppose that the below part of the Penrose diagram divided by shock waves is a black hole geometry with mass $M - E$ and that the left and right parts are mass $M$ as a result of inserting energy $E$. See Figure the left figure in figure \ref{fig:AdSdshock}.
\ \\
The DTR condition \eqref{DTRcondition} at the intersection radius $r=r_c$ becomes  
\begin{align}
\label{solDTRAdS}
(r_c^2-8G_N \ell^2 M_t)(r_c^2-8G_N \ell^2 (M - E)) &=(r_c^2-8G_N \ell^2 M)^2
\end{align}
which determines the relation between $M_t$ and $r_c$ as 
\begin{align}
\label{MtMErcrela}
M_t =M+E+\frac{8G_N \ell^2 E^2}{8G_N  (M - E) \ell^2 -r_c^2}
\end{align}
In the right figure \ref{fig:AdSdshock}, simply the role of above and below patches is flipped.\\
{\it \underline{Consistency check}}\\
The left/right patch and the below patch have masses $M$ and $M-E$, respectively in the left figure \ref{fig:AdSdshock}. 
This mass difference is the same as the single shock wave case.  Therefore, the continuity of $r$ across the shock wave reproduces the constant shift of $\alpha$ given in \eqref{UtildeUrela} in the double scaling limit eq.~\eqref{BTZDS}.

Next, let us look at the relationship between the left (or right) and the above patch in the left figure \ref{fig:AdSdshock}. In the double scaling limit eq.~\eqref{BTZDS}, 
\begin{align}
r^2_c \to R^2 = 8G_N \ell^2 M \,, \quad G_N E \to 0  \,,
\end{align} 
then, the denominator of the last term in \eqref{MtMErcrela} becomes zero and we need careful analysis in that limit. 

To understand $M_t$ in detail, let us first determine $r_c$ as a function of $t_w$. Since 
both shock waves are inserted at $t=t_w$ on the left and right boundary, we have 
\begin{align}
U = V = \pm e^{- \frac{R t_w}{\ell^2}} 
\end{align} 
at the intersection $r =r_c$. Here again, the positive and negative signs represent the case in the left and right figures of figure \ref{fig:AdSdshock}, respectively. 

Therefore, we have 
\begin{align}
U V = e^{-2 R t_w/\ell^2}=e^{2 R r^{\ast}(r_c)/\ell^2}=\frac{R-r_c}{R +r_c}
\end{align}
which determines $t_w$ dependence of $r_c$ as 
\begin{align} 
r_c= R\tanh \frac{R}{\ell^2}t_w
\end{align}
From this, $M_t$ in \eqref{MtMErcrela} becomes
\begin{align}
\label{MtMEcoshrela}
M_t&=M+E+\frac{E^2\cosh ^2\frac{R}{\ell^2}t_w}{M-E\cosh ^2\frac{R}{\ell^2}t_w} \,.
\end{align}

Let us consider the double scaling limit \eqref{BTZDS}.  
which is $E/M \to 0$ with $t_w \to \infty$ where $\alpha$ is fixed. 
In this limit 
\begin{align}
M \ll E \cosh^2 \frac{R}{\ell^2} t_w
\end{align}
and therefore $M_t$ becomes 
\begin{align}
M_t \to M   
\end{align}
Therefore, the effects of the shock waves are restricted only on the horizon $U = 0$ and $ V =0$ in the double scaling limit \eqref{BTZDS} which is consistent with the metric eq.~\eqref{twodeltashock}.

%

\subsection{de Sitter spacetime}
\label{sec:shockgeodS}

We now consider the case of dS. The method is the same as for the BTZ case. However, the results are different in a very interesting way.

The (Schwarzschild) dS spacetime \cite{Deser:1983nh, Spradlin:2001pw} can be written as
\begin{align}
&ds^2=-f(r)dt^2+\frac{dr^2}{f(r)}+r^2 d\phi^2,\\
&f(r)= \frac{L^2-r^2}{\ell^2}.
\end{align}
where $\ell$ is dS-radius. $L$ is a parameter for the cosmological horizon, such that \begin{align}
L^2 & =\ell^2 \quad \quad \quad \quad \quad \quad \quad  \mbox{(for pure dS)}  \\
L^2 & = (1-8G_N M)\ell^2 \quad \, \mbox{(for Schwarzschild dS)}
\end{align} 
Note that even though we call Schwarzschild dS (or SdS in short), only a cosmological horizon exists and there is no black hole horizon in three dimensions. 

Let us rewrite this in Kruskal coordinates defined by
\begin{align}
\label{KruskalfordS}
&U_R= e^{\frac{L}{\ell^2}(t_R-r^{\ast})},\ V_R=-  e^{-\frac{L}{\ell^2}(t_R+r^{\ast})},
\end{align}
where $r^{\ast}(r)$ is a tortoise coordinate,
\begin{align}
\label{tortoisedS}
&r^{\ast}(r)=\int \frac{dr}{f(r)}=\frac{\ell^2}{2L}\log \left| \frac{L+r}{L-r}\right| \,.
\end{align}
at the cosmological horizon $r \to L$, $r^{\ast}(r) \to + \infty$. 
Note that the sign in eq.~\eqref{KruskalfordS} is different from that of AdS in eq~\eqref{KruskalRBTZ}. 

Again the subscript $R$ in $t_R$ indicates that this definition only covers the right static patch. 
We sometimes call the right static patch $r=0$ a south pole and the left static patch $r=0$ a north pole.  
In this coordinate, the dS metric is
\begin{align}
ds^2=\frac{-4\ell^2 dUdV}{(1-UV)^2}+L^2 \left( \frac{1+UV}{1-UV} \right)^2 d\phi^2
\end{align}
and the cosmological horizon is $U=0$ or $V=0$. 

It is also useful to define the null coordinates $(u, v)$ as follows,  
\begin{align}
\label{dSuvcoord}
u_R=t-r^{\ast},\ v_R=t+r^{\ast} \,, \quad  u_L=t-r^{\ast},\ v_L=t+r^{\ast} 
\end{align}
Note that this definition of $(u,v)$ is the same as that of AdS in eq.~\eqref{uvRforAdS}, however, the orientations of the $u$ and $v$ coordinates are switched compared to the AdS case, and so are $(U, V)$ coordinates. See figure \ref{fig:dS-coordinate} for the orientations of the coordinates. 
Note that in the right static patch, the south pole,  $r=0$, corresponds to $r^{\ast} = 0$, and there, $u_R = v_R = t$ ranges from $-\infty$ to $\infty$.

The ingoing Eddington-Finkelstein coordinates are   
\begin{align}
\label{dSuvmet}
ds^2 
&=-f(r)du_R^2 -2du_Rdr+r^2 d\phi^2.
\end{align}
where $(u_R, r)$ coordinates the right static patch and future patch which is $ r \ge L$. 

\subsubsection{Single shock wave}
{\it Method} 1\\
As is the case of BTZ, we can find the solution of the Einstein equation under the stress tensor
\begin{align}
\label{stressfordS3}
T_{UU}=\frac{\beta}{4\pi G_N }\delta(U)
\end{align}
where $\beta > 0$ due to ANEC.   For simplicity, 
we are considering a $2+1$ dimensional solution, but the solutions in higher dimension were specifically constructed in \cite{Hotta:1992qy,PhysRevD.47.3323,Sfetsos:1994xa}. These solutions can be obtained by appropriately boosting the solution for a point particle. 

\begin{figure}[h]
\centering
\includegraphics[keepaspectratio,scale=0.25]{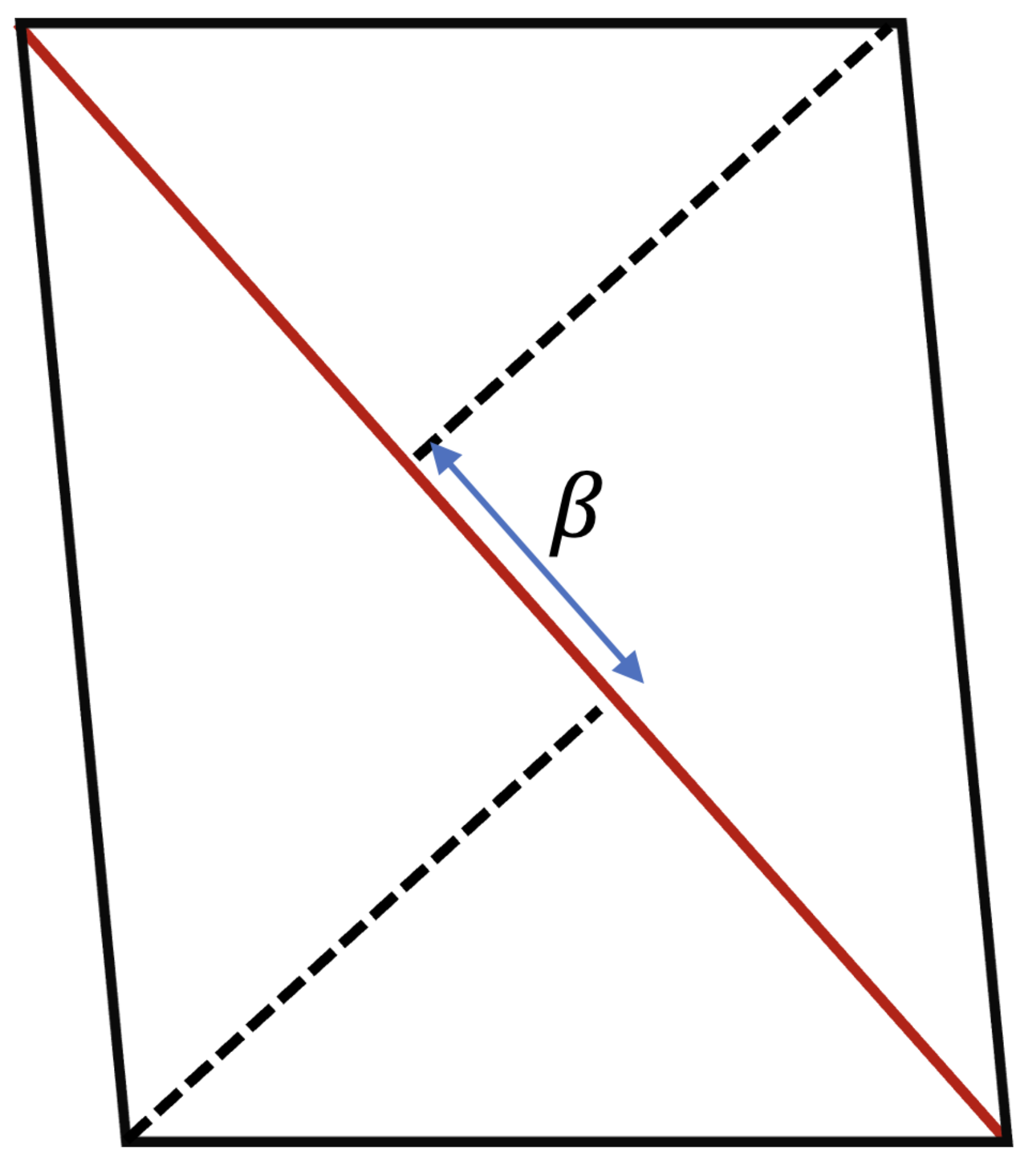}
\caption{Penrose diagram of dS with single shock wave which is inserted on the cosmological horizon. The shift is exactly the opposite compared with the BTZ case as Figure \ref{fig:AdSshockPen}. In this case, effectively a time advance of $\beta$ is generated. }
\label{dSshockPen}
\end{figure}

\begin{figure}[h] 
\centering
\includegraphics[keepaspectratio,scale=0.42]{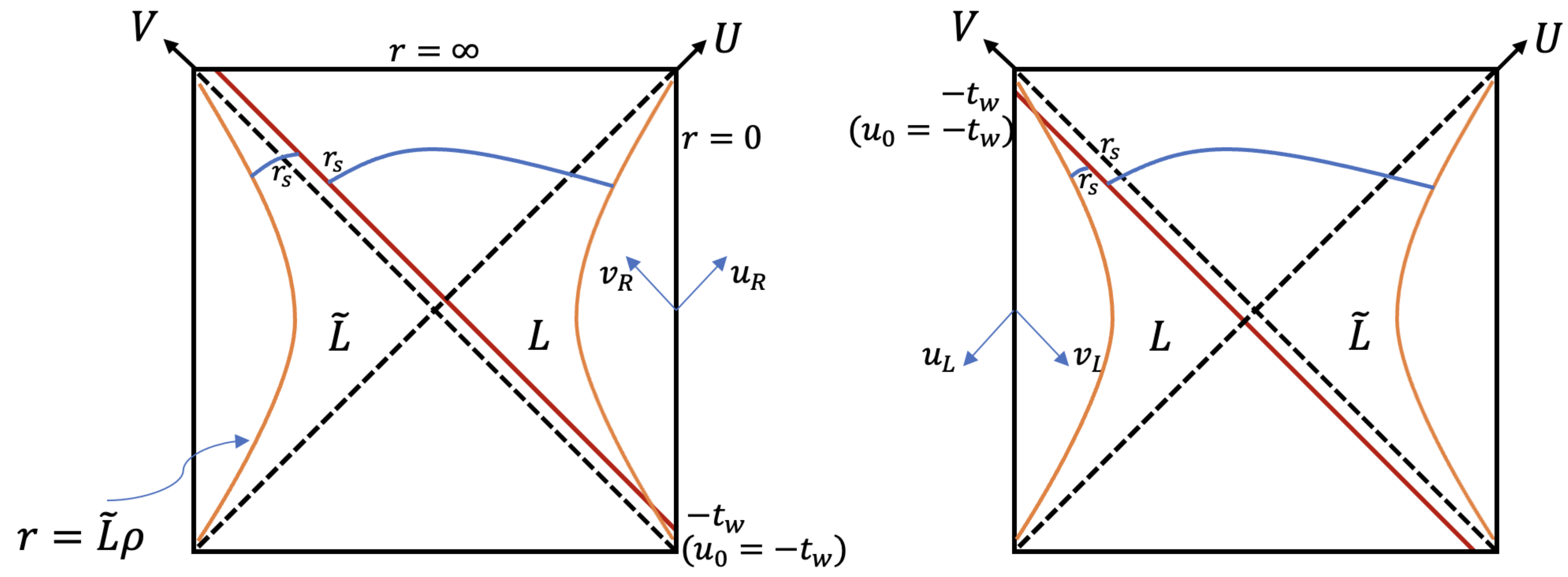}
\caption{Penrose diagram of single shock wave on dS geometry. Note that our convention of  $U$ and $V$ orientation is switched compared with the BTZ case as Figure \ref{fig:AdS-coordinate}. We treat the case where a single shock wave is inserted at $t_R = -t_w$ ($t_L = -t_w$) in the south pole (north pole). 
The stretched horizon is drawn in the orange curve, which is $r = \rho L$ and $\rho \to 1$. $L$ and $\tilde{L}$ are shifted by the inserted energy $E$. Note, however, that the relationship is inverse compared with the BTZ case. For example, if $L$ is for pure dS, then $\tilde{L}$ is for SdS. $r$ is again continuously connected in this case.}
\label{fig:dS-coordinate}
\end{figure}

The Einstein equation with stress tensor eq.~\eqref{stressfordS3} yields 
\begin{align}
\label{dSsingleshockdelta}
ds^2=\frac{-4\ell^2 dUdV}{(1-UV)^2}-4 \ell^2\beta  \delta(U)dU^2+L^2 \left( \frac{1+UV}{1-UV} \right)^2 d\phi^2 
\end{align}
In this metric, the effects of the shock wave are restricted only on the horizon $U=0$.  
It is convenient to rewrite this metric as follows.
\begin{align}
\label{shockgeo2}
ds^2=\frac{-4\ell^2 dUdV}{(1-U(V- \beta \theta(U)))^2}
+L^2 \left( \frac{1 +U(V- \beta \theta(U))}{1-U(V-  \beta \theta(U))} \right)^2 d\phi^2
\end{align}
Setting $V \to V+ \beta \theta(U)$, eq.~\eqref{shockgeo2} becomes \eqref{dSsingleshockdelta}. 
Note that compared with the BTZ case in eq.~\eqref{shockgeo12} and \eqref{shockgeo1}, the effect of the shock wave is exactly the {\it opposite}. This reverse shift behaves as if it induces a time advance. However, this is causally consistent \cite{Gao:2000ga, Aalsma:2020aib}. Shock wave makes the universe shrink gravitationally and allows a causal connection between the left and right patch. See figure \ref{dSshockPen}.
\\
{\it Method} 2\\
Now let us see the second method by preparing Schwarzschild dS spacetime with different masses and connecting them at the shock wave. In the case of dS, the mass shift is nontrivial, so we will use $(U, V)$ for the upper part and $(\tilde{U},\tilde{V})$ for the lower part as in Penrose diagram in the left figure of figure \ref{fig:dS-coordinate}. Again, in the right figure, the role of $U$ and $\tilde{U}$ is reversed. 
The connection condition is the same as BTZ, making $r$ continuous on the shock wave, 
\begin{align}
\label{rcontinuityfordS}
L \frac{1+UV}{1-UV}=\tilde{L} \frac{1+\tilde{U}\tilde{V}}{1-\tilde{U}\tilde{V}}
\end{align}
Thus, the metric must be $C^0$ and it can also be written in the Vaidya metric as 
\begin{align}
\label{Vaidya2dSLLt}
ds^2=-\frac{ \left( L^2 \theta(u-u_0) + \tilde{L}^2 \theta(u_0- u) - r^2 \right) }{\ell^2} du^2-2du dr+r^2d\phi^2
\end{align}
where $u$ implies $u_R$ ($u_L$) in the left (right) figure in Figure \ref{fig:dS-coordinate}. $u_0$ is a parameter for the shock wave trajectory; the shock wave is on the constant surface 
\begin{align}
u_{{R(L)}}=u_0 = - t_w
\end{align} 
in $uv$ coordinates, and $t=-t_w$ is the time on the south (north) pole, {\it i.e.,} right (left) patch $r=0$, and   
\begin{align}
\label{dSshockwaveU}
U= \pm e^{- \frac{L}{\ell^2} t_w} \,, \quad \tilde{U}=  \pm e^{- \frac{\tilde{L}}{\ell^2} t_w}
\end{align} 
where we take the positive (negative) sign for the case in the left (right) figure in figure \ref{fig:dS-coordinate}.

This metric is the solution of Einstein equation with the stress tensor 
\begin{align}
T_{uu} = \frac{L^2 - \tilde{L}^2}{16 \pi G_N \ell^2 r} \delta(u-u_0)
\end{align}
Setting 
\begin{align}
\label{dSLtildeLrela}
L^2 = \ell^2 - 8 G_N M \ell^2 \,, \quad \tilde{L}^2 = \ell^2 - 8 G_N \tilde{M} \ell^2
\end{align}
we obtain 
\begin{align}
\label{TuufordS}
T_{uu} = \frac{\tilde{M} - M}{2 \pi  r}\delta(u-u_0) \equiv  \frac{E}{2 \pi  r}\delta(u-u_0) \,.
\end{align}
Therefore the ANEC requires 
\begin{align}
\tilde{M} > M \, \Leftrightarrow E > 0  \,.
\end{align}
Thus, after the shock wave, the mass should {\it decrease} in the dS case.  For example, if $M=0$, the $L$ is for the dS cosmological horizon. Then $\tilde{M} > 0$, where $\tilde{L}$ is for SdS. 
\\
%
{\it \underline{Consistency check}}\\
As is the case of AdS,  we also check if {\it Method} 1 and {\it Method} 2 are consistent in the case of dS, and  
for that purpose, we need the following double scaling limit
\begin{align}
\label{DSdS}
G_N  E \to 0 \,, \quad t_w \to \infty  \quad \mbox{where}  \quad G_N E \, e^{\frac{L}{\ell^2}t_w} = \mbox{fixed} \,.
\end{align}
Under the double scaling limit eq.~\eqref{DSdS}, we can solve $r$-coordinate continuity condition eq.~\eqref{rcontinuityfordS} for $\tilde{U}$, with \eqref{dSshockwaveU} and \eqref{dSLtildeLrela} and we obtain $\beta$  as  
\begin{align}
\label{dSbeta}
\tilde{V} = V  \pm \beta \,, \quad 
\beta = \frac{2 G_N E}{{1 - 8 G_N M}}  \, e^{\frac{L t_w}{\ell^2}}  
=  \frac{2 G_N E \ell^2}{L^2}  \, e^{\frac{L t_w}{\ell^2}}  > 0 \,.
\end{align}
Again, the positive (negative) sign for the left (right) figure in figure \ref{fig:dS-coordinate}. 
Since $L$ and $\tilde{L}$ are switched between the left and right figure, 
eq.~\eqref{dSbeta} is exactly what we obtained in the metric eq.~\eqref{shockgeo2}.

The other thing to check is that the stress tensor of the Vaidya metric, which describes the mass shift in an obvious way, does indeed match that of {\it Method} 1.
One can also check the stress tensor \eqref{TuufordS} matches with \eqref{stressfordS3} in the double scaling limit \eqref{DSdS}, where $r \to L$ as follows
\begin{align}
T_{UU} =  \left(\frac{du}{dU}\right)^2 T_{uu} \to
\left(\frac{du}{dU}\right) \frac{E}{2 \pi L} \delta(U)
=\frac{\beta}{4\pi G_N }\delta(U) \quad \mbox{at } r\to L
\end{align}
where $\beta$ is given by \eqref{dSbeta}. 
This is indeed consistent with the stress tensor in \eqref{stressfordS3}.

Before we go on to double shock waves, we summarize the results and emphasize the difference between BTZ and dS. 
First of all, the shift of coordinate translation at the shock wave in dS is the exact opposite of that in the BTZ, which allows a time advance. In this way, it is possible to travel from the south pole to the north pole. Second, the mass shift is also the exact opposite of the BTZ. In other words, in the Penrose diagram, the lower part is the Schwarzschild dS spacetime which has a larger mass compared with the upper part. For example, the lower part is the Schwarzschild dS spacetime and the upper part can be pure de Sitter spacetime, see the left figure \ref{fig:dS-coordinate}.

\subsubsection{Double shock waves}
{\it Method} 1\\
As in the case of AdS, we consider following stress tensors 
\begin{align}
T_{VV}=\frac{\beta}{4\pi G_N }\delta(V) \,, \quad T_{UU}=\frac{\beta}{4\pi G_N } \delta(U) \,.
\end{align}
which yields the solution as follows
\begin{align}
\label{doubleshockbeta}
ds^2=\frac{-4\ell^2 dUdV}{(1-UV)^2}-4 \ell^2\beta  \delta(U)dU^2 -4 \ell^2\beta  \delta(V)dV^2  +L^2 \left( \frac{1+UV}{1-UV} \right)^2 d\phi^2 
\end{align}
Again this is the solution only in the leading order of $\beta$, neglecting $O(\beta^2)$. Therefore we assume $\beta$ is small. As in the BTZ case, this should be a shift solution similar to the single shock wave case if we consider $V\neq 0$ or $U\neq 0$. \\
\begin{figure}[h]
\centering
\includegraphics[keepaspectratio,scale=0.42]{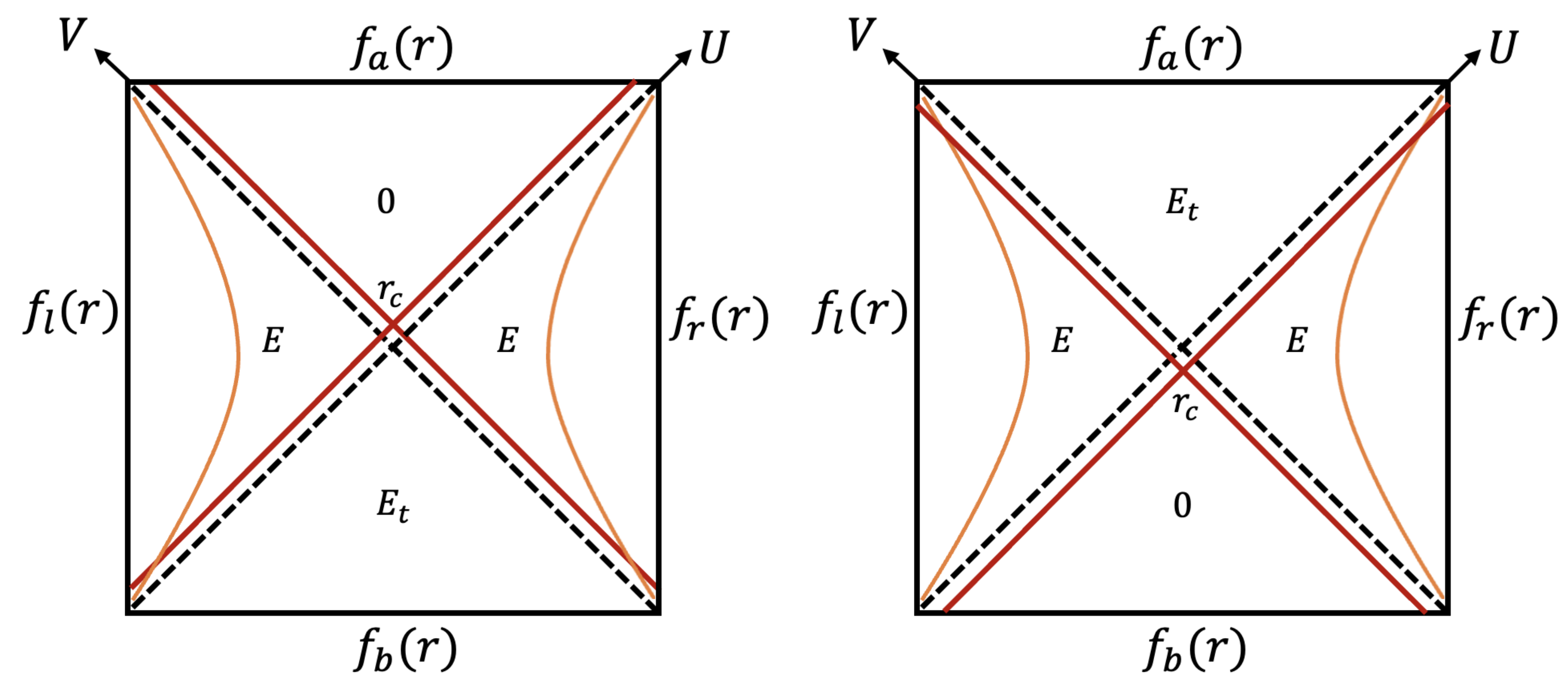}
\caption{dS space-time divided by two shock waves (red solid lines). For simplicity, we consider the insertion of symmetric double shock waves. Again there are two choices for the shock waves, as left and right figures. 
On the left (right) figure, the shock wave is inserted $t_R = -t_w$ ($t_L = - t_w$). Another shock wave is determined such that the Penrose diagram is symmetric between $l$(eft) and $r$(ight) as these two figures.  
Shock waves divide the spacetime into four distinguished region as $a$(bove), $b$(elow), $l$(eft), $r$(ight), and their masses are $0$, $E_t$, $E$ and $E$ respectively in the left figure. Two shock waves intersect at $r=r_c$.}
\label{fig:dSdshock}
\end{figure}
\ \\
{\it Method} 2\\
In the case of dS with double shock waves, we need to consider the DTR condition. We consider the case where one shock wave is inserted from the past on the south (or north) pole at $t_R = -t_w$ ($t_L = - t_w$) in the left (right) figure in figure \ref{fig:dSdshock}. The other shock wave position is automatically determined so that the Penrose diagram is symmetrical between left and right. 

For simplicity, let us consider the setting of the left figure \ref{fig:dSdshock}, and we take the above part of the shock wave as simply pure dS, and set the mass of Schwarzschild de Sitter in the below part as $E_t$, then 
\begin{align}
&f_a(r)=1-\frac{r^2}{\ell^2},\\
&f_l(r)=f_r(r)=1-8G_N E-\frac{r^2}{\ell^2},\\
&f_b(r)=1-8G_N E_t-\frac{r^2}{\ell^2}.
\end{align}
In this case, the DTR condition becomes, 
\begin{align}
\label{DTREq}
\left(1-\frac{r_c^2}{\ell^2}\right)\left(1-8G_N E_t-\frac{r_c^2}{\ell^2}\right)=\left(1-8G_N E-\frac{r_c^2}{\ell^2}\right)^2
\end{align}
at the shock wave intersection $r=r_c$. 

In the case of the right figure \ref{fig:dSdshock}, the a(bove) and b(elow) parts are exchanged, but the DTR condition is the same.  

Therefore in both left and right figures in figure \ref{fig:dSdshock}, we can calculate $E_t$
\begin{align}
\label{EtErcrela}
E_t=2E+\frac{8G_N \ell^2 E^2}{r_c^2-\ell^2} \,.
\end{align}
Here $r_c$ is 
\begin{align}
\label{rctwdS}
r_c=\frac{\tilde{L}}{\tanh \frac{\tilde{L} t_w}{\ell^2}}
\end{align}
with 
\begin{align}
\label{Ltildedef}
\tilde{L}^2 = \ell^2 \left( 1 - 8 G_N E \right)  \,.
\end{align} 
This can be seen since the two shock waves intersect at 
\begin{align}
U = V = \pm e^{- \frac{\tilde{L} t_w}{\ell^2}} 
\end{align}
where the positive (negative) sign is for the left (right) figure in figure \ref{fig:dSdshock}. Then at the shock wave intersect, 
\begin{align}
U V = e^{- \frac{2 \tilde{L} t_w}{\ell^2}} =e^{- \frac{2 \tilde{L} r^{\ast}(r_c)}{\ell^2}} = \frac{r_c - \tilde{L}}{r_c + \tilde{L}} \,,
\end{align}
which gives eq.~\eqref{rctwdS}. 

Pluggin eq.~\eqref{rctwdS} and \eqref{Ltildedef} into \eqref{EtErcrela}, we obtain  
\begin{align}
\label{EtErcrela2}
E_t=2E+\frac{8G_N \ell^2 E^2}{r_c^2-\ell^2} \,.
\end{align}
\\
{\it \underline{Consistency check}}\\
In the double scaling limit eq.~\eqref{DSdS}, 
$\tilde{L} \to \ell$ and 
\eqref{EtErcrela2} goes to 
\begin{align}
G_N E_t& \to  
2 G_N^2 E^2 e^{\frac{2 t_w}{\ell}}  = \frac{\beta^2}{2}  \ll 1
\end{align}
which is $O(\beta^2)$ and negligible, where $\beta$ is given by eq.~\eqref{dSbeta} with $L = \ell$. This is consistent since eq.~\eqref{doubleshockbeta} is a solution neglecting this order.   
%
%
%
\section{Delay of Hyperfast}\label{sec:delay}
\subsection{Complexity and time dependence of WdW patch}
Quantum complexity $\mathcal{C}(| \psi \ket)$ for the target state $| \psi \ket$ is 
the minimum number of elementary gates required to build that state from a reference one.  We can consider the case where the target state $| \psi \ket$ is 
an entangled state of two boundary theories, that live on the left and right boundaries. 
Given the target state $| \psi \ket$, which is a function of both the left and right 
boundary time $t_l$, $t_r$, then $\Sigma$ is a codimension one bulk Cauchy surface anchored at the boundary state $t_l$ and $t_r$.  
These are three main holographic proposals that are dual to quantum complexity $\mathcal{C}[| \psi \ket]$.  
\\
{\bf CA (Complexity=Action)}\\
This is the gravitational bulk action of the Wheeler-de Witt (WdW) patch of bulk causal diamond of Cauchy slice $\Sigma$, 
\begin{align}
\mathcal{C}_A(\Sigma)=\frac{I_{\rm WdW}}{\pi}
\end{align}
{\bf CV2.0}\\
This conjecture is similar to CA, but the difference is in CV2.0, the holographic Complexity is the space-time volume of the WdW patch
\begin{align}
\mathcal{C}_W(\Sigma)=\frac{V_{\rm WdW}}{G_N \ell_{\rm bulk}^2}
\end{align}
instead of the action. In pure dS and pure AdS case, Lagrangian density is constant and proportional to $\ell_{\rm bulk}^{-2}$, therefore both CA and CV2.0 behave in similar ways\footnote{Precisely speaking, the behavior is generally different because of the boundary terms. However, the critical time $\tau_{\infty}$ given in \eqref{behaviornearc} is determined by the WdW patch in both CA and CV2.0, as examined by \cite{Jorstad:2022mls}.}.
Here $\ell_{\rm bulk}$ is the length scale set by the cosmological constant. \\
{\bf CV (Complexity=Volume)}\\
In this conjecture, the Complexity is evaluated by the maximal volume of codimension one surface $\Sigma$, 
\begin{align}
\mathcal{C}_V=\max  
\left[\frac{V({\Sigma})}{G_N \ell_{\rm bulk}}\right]
\end{align}
The maximization should be conducted under the condition that the boundary time $t_l$ and $t_r$ are fixed. 
\\
Using all these three methods, the complexity calculations in general dS$_{d+1}$ ($d \ge 2$) were made in \cite{Jorstad:2022mls}. As a result, they found that complexity and its growth rate diverge when a certain critical time $\tau_{\infty}$ is reached, no matter which conjecture is followed, 
\begin{align}
\label{behaviornearc}
\lim_{\tau \rightarrow \tau_{\infty}}\mathcal{C} \to \infty,\ \ \lim_{\tau \rightarrow \tau_{\infty}}\frac{d \mathcal{C}}{d\tau} \to \infty.
\end{align}
This is typical hyperfast behavior in dS$_{d+1}$ ($d \ge 2$). 
The main contribution to this divergence is $r_f$ reaches infinity, where $r_f$ is the radius of the conical point at the most future tip of the WdW patch. In other words,  the fact that the radius $r = r_f$ at the top vertex of the diamond-shaped-WdW patch diverges causes the volume of $S^{d-1}$ to diverge, and as a result, causes the complexity diverges. 
In the case of CV, $r = r_{\rm turn}$ diverges, which we will define later but $r_{\rm turn}$ is the radius of most future tip in $\Sigma$ slice.  
See Figure \ref{fig:dS-Comp1} and \ref{fig:dS-CVarg}. 

In particular, let us look at the time dependence of the WdW patch. 
Note here that the holographic screen is on the stretched horizon $r=\rho\, \ell$, which is close to the cosmological horizon. Here $r=\ell$ is the cosmological horizon for pure dS and $\rho$ is a 
parameter for the stretched horizon, {\it i.e.,} $0<\rho <1$ and $\rho \to 1$\footnote{The value of $\rho$ is not strongly restricted in the following analysis. $\rho\in (0,1)$, which should be neither strictly 0 nor strictly 1. Note, however, that according to Susskind’s conjecture \cite{Susskind:2021esx}, it is actually preferable for $\rho$ to be as close to 1 such that de Sitter RT formula \cite{Ryu:2006bv} to work. }. 
We choose the boundary time increases upward on both the left and right stretched horizons such that $t_R=-t_L\propto \tau$ on the stretched horizon. As $\tau$ increase, 
the radius $r$ at the top vertex of the diamond-shaped-WdW patch also increases. 
Then at some finite time, $r_f$ will eventually reach infinity, and then, at the moment of arrival, the volume of the WdW patch also grows rapidly and diverges in dS$_{d+1}$ ($d \ge 2$). Therefore, the critical time can be estimated as the time when $r_f=\infty$. 

The above arguments about hyperfast are basically valid for dimensions greater than two. In two dimensional JT gravity, we can see the same behavior in CA. However, CV and CV2.0 in JT gravity show different behavior. The volume of the geodesic line does not diverge even just at the critical time and it is  $O(1)$ \cite{Chapman:2021eyy}. 
This is because we ignore the effect of dilaton. Note that CA conjecture takes into account dilaton but both CV and CV2.0 ignores the dilaton.   
The divergence of the volume of $S^{d-1}$ in $d\ge 2$ is reflected as the dilaton divergence in JT gravity. Therefore with CA conjecture, it shows the divergence at the critical time  \cite{Anegawa:2023wrk}.

We will now study how this critical time is affected by shock waves. Even in the absence of a shock wave, the critical time can be calculated by the time when $r_f$ becomes infinite \cite{Susskind:2021esx, Jorstad:2022mls}. However, the shock waves affect the critical time because the shape of the WdW patch is modified. We are particularly interested in, by the shock wave perturbation, how the dS complexity critical time $\tau_{\infty}$ is modified, especially whether $\tau_{infty}$ becomes greater or lesser by shock waves. 
Our argument is based on both the WdW patch, which is for CV2.0 and CA, and also the complexity = volume (CV) method. We will find in both methods, the results are the same and therefore consistent. 

%
\subsection{BTZ case}

\begin{figure}[h]
\centering
\includegraphics[keepaspectratio,scale=0.45]{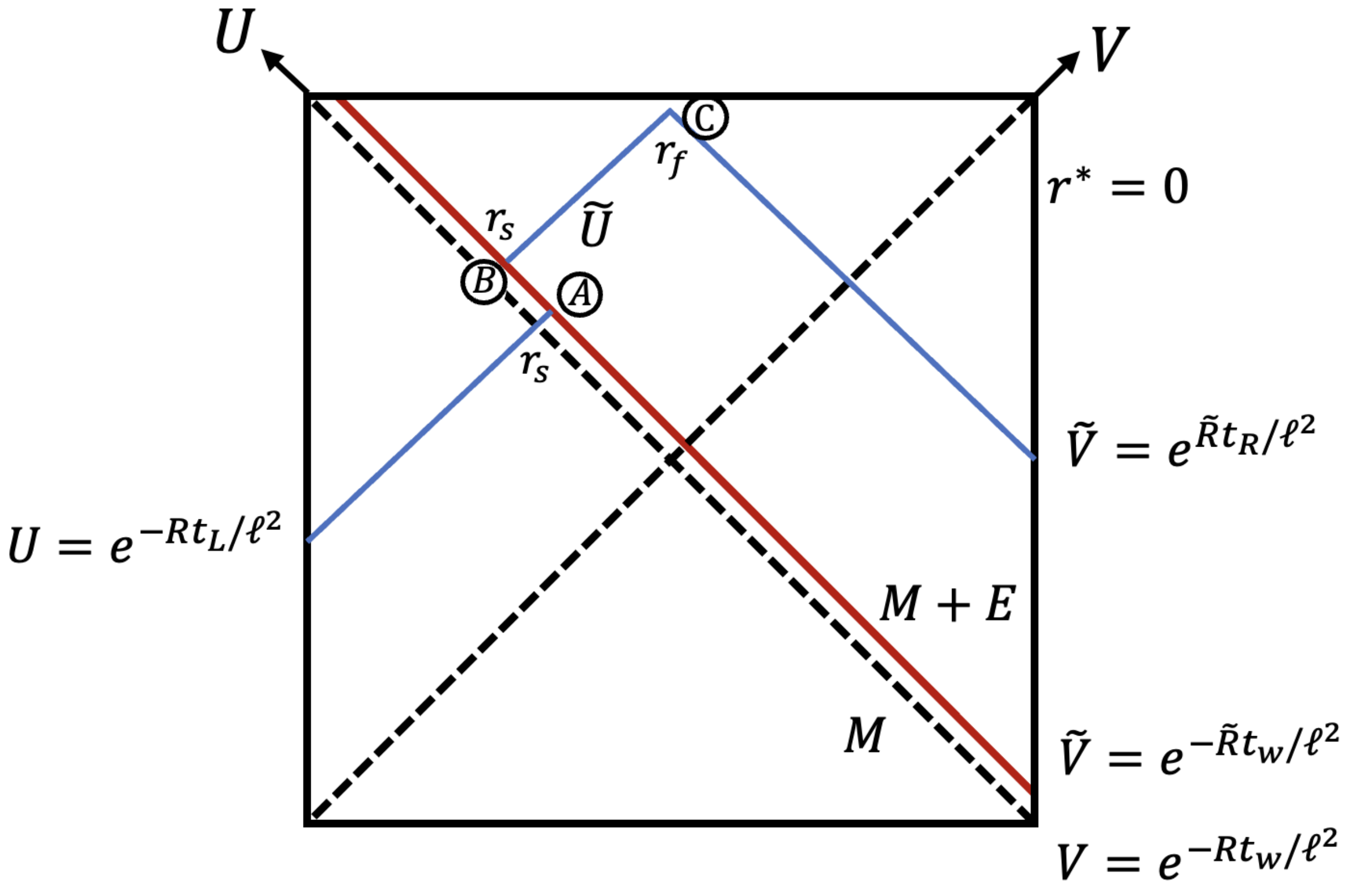}
\caption{The future null boundaries of the WdW patch are drawn in BTZ geometry. 
We use $(\tilde{U}, \tilde{V})$/$({U}, {V})$ after/before the shock wave (red line). 
$r_f$ is the radius of conical point where fixed $\tilde{U}$ and fixed $\tilde{V}$ of the null boundaries (blue line) meet. Point A and B have the same radius $r=r_s$, but $U$ is shifted by the shock wave.
The ``critical time'' is the time at the boundary where $r_f \to 0$.}
\label{fig:AdS-Comp1}
\end{figure}

Before we proceed to the dS case, let us calculate the critical time in the BTZ case as a warm-up.\\
{\bf Single shock wave}\\
Given the boundary time $t_L$ and $t_R$, the WdW patch extends from there. See Figure \ref{fig:AdS-Comp1}, where the future null boundaries of the WdW patch are drawn. 
The value of $U$ or $V$ on each null boundary is determined by the boundary time. Note that $t_L$ goes backward on the left boundary. However, due to the shock wave, the value of $U$ is not continuous. To distinguish before and after the shock wave, we use $(\tilde{U}, \tilde{V})$ coordinate after the shock wave and $({U}, {V})$ before the shock wave. 

As Figure \ref{fig:AdS-Comp1}, let us define the $r_f$ as the radius where the $U=$ constant and $V=$ constant null lines meet, namely the top ``tip'' of WdW patch where the radius $r$ becomes the smallest.
Even in the BTZ case, we can define the “critical time” as the boundary time when the intersection point, $r_f$, reaches the singularity, {\it i.e.,} $r_f = 0$. In BTZ, this does not give a significant contribution, since the volume near the singularity $r=0$ vanishes\footnote{Because of this, although $r=0$ is a region where the semiclassical approximation inherently breaks down, formally one can compute the critical time. More specifically, even if we introduce an appropriate cutoff such as  $r > \epsilon$, the main contribution to the Complexity is the volume away from the small $r$ region.}. However, the shape of the WdW patch will change after this time.

Under this setting, we can write down the relationship between $r_f$ and 
$t_w$, the boundary time shock wave is inserted, and $t_R$ and $t_L$. 
It is also convenient to define $r_s$ as a radius where the shock wave and null line of fixed $U$ from the left time $t_L$, intersect, denoted as point A in Figure \ref{fig:AdS-Comp1}.

When $U=$ constant and $V=$ constant null lines intersect at a certain radius $r$, the following holds, 
\begin{align}
\label{UVandrcond}
UV &= \pm e^{\frac{2R r^{\ast}(r)}{\ell^2}} \,, \\
\tilde{U} \tilde{V} &=  \pm e^{\frac{2\tilde{R} \tilde{r}^{\ast}(r)}{\ell^2}}
\label{UVandrcond2}
\end{align}
where $ \tilde{r}^{\ast}(r)$ is a tortoise coordinate as eq.~\eqref{tortoiseAdS} but in terms of $\tilde{R}$ instead of $R$. 
For $\pm$, we choose $-$ if $r > R$ and $+$ if $r < R$. 
Eq.~\eqref{UVandrcond} is valid in $(U, V)$ coordinate patch, and \eqref{UVandrcond2} is in $(\tilde{U}, \tilde{V})$ coordinate patch.
Note that the shock wave is inserted at $t_R=-t_w$, which is the boundary between $(U, V)$ and  $(\tilde{U}, \tilde{V})$ coordinate patch. Therefore, there are two definitions of $V$ here, one using $R$ and the other using $\tilde{R}$ as 
\begin{align}
\label{singleshockBTZ}
\qquad \qquad V=e^{- \frac{Rt_w}{\ell^2}}  \,, \quad \tilde{V}=e^{- \frac{\tilde{R}t_w}{\ell^2}}  \quad \mbox{(on the shock wave)}
\end{align}

Before proceeding to the evaluation of the critical time, we first confirm a constant shift occurs between $U$ and $\tilde{U}$ in the double scaling limit \eqref{BTZDS}. 
Since $U$ = constant and $\tilde{U}$ =constant intersect with the shock wave given by eq.~\eqref{singleshockBTZ} at $r=r_s$, using \eqref{UVandrcond} and \eqref{UVandrcond2}, 
we have 
\begin{align}
\label{shiftfromr}
\tilde{U}-U= e^{ \frac{ \tilde{R}(t_w+2 \tilde{r}^{\ast}(r_s))}{\ell^2}}-e^{\frac{ R(t_w+2 r^{\ast}(r_s))}{\ell^2}}
\end{align}
Substituting $\tilde{R}=\sqrt{\frac{M+E}{M}}R$, using the fact that in the double scaling limit \eqref{BTZDS}, $r_s \to R$, the right hand side of \eqref{shiftfromr} becomes 
\begin{align}
\label{UtildeUrela2}
\tilde{U}-U= e^{ \frac{ \tilde{R}(t_w+2 \tilde{r}^{\ast}(r_s))}{\ell^2}}-e^{\frac{ R(t_w+2 r^{\ast}(r_s))}{\ell^2}} \to \frac{E}{4M}e^{\frac{Rt_w}{\ell^2}}=\alpha
\end{align}
which is exactly what we have seen in \eqref{UtildeUrela}.

Now, let's calculate the critical time $\tau_\infty$.  It must be clear from Figure \ref{fig:AdS-Comp1} that the critical time is smaller than that without a shock wave. In particular, since critical time is zero without a shock wave, it must be negative.  
In order to evaluate the time shift,  using eq.~\eqref{UVandrcond} for point A and \eqref{UVandrcond2} for point B and C, 
\begin{align}
\label{UVcondA}
e^{-\frac{R t_L}{\ell^2}} e^{-\frac{Rt_w}{\ell^2}} &=e^{\frac{ 2R r^{\ast}(r_s)}{\ell^2}} \quad \mbox{(point A)}\\
\label{UVcondB}
\tilde{U}e^{- \frac{ \tilde{R}t_w}{\ell^2}} &=e^{\frac{ 2\tilde{R} \tilde{r}^{\ast}(r_s)}{\ell^2}} \quad \mbox{(point B)}\\
\label{UVcondC} 
\tilde{U}e^{\frac{ \tilde{R}t_R}{\ell^2}} &=e^{\frac{2\tilde{R} \tilde{r}^{\ast}(r_f)}{\ell^2}} \hspace{-0.3mm}\quad \mbox{(point C)}
\end{align}
where $\tilde{U}$ is the coordinate value on the null line between points B and C, and we use 
\begin{align}
U =e^{-\frac{R t_L}{\ell^2}} \,, \quad \tilde{V} = e^{\frac{ \tilde{R}t_R}{\ell^2}}
\end{align}
as shown in figure \ref{fig:AdS-Comp1}.

Eq.~\eqref{UVcondA}, and the equation \eqref{UVcondC} divided by \eqref{UVcondB} yield,  
\begin{align}
\label{UVcondsing2}
t_L+t_w &=-2r^{\ast}(r_s) \,, \\
t_R+t_w &=2  \left( \tilde{r}^{\ast}(r_f)- \tilde{r}^{\ast}(r_s) \right)
\end{align}
respectively. 
Eliminating $t_w$ from the above and setting $-t_L=t_R \equiv \tau$, 
we have 
\begin{align}
2 \tau = t_R - t_L = 2 \left(  \tilde{r}^{\ast}(r_f) + r^{\ast}(r_s)   - \tilde{r}^{\ast}(r_s) \right) 
\end{align}
Since at $r_f \to 0$, $ \tilde{r}^{\ast}(r_f) \to 0$,  this leads to the following as the advanced part of critical time.
\begin{align}
\label{criticalad1}
 \Delta \tau_{\infty}= \tau_{\infty}   =r^{\ast}(r_s)- \tilde{r}^{\ast}(r_s) 
\end{align}
$ \Delta \tau_{\infty}= \tau_{\infty} $ because critical time without shock wave is at $\tau =0$. 

To see $\tau_{\infty} < 0$, we need to evaluate the right-hand side of eq.~\eqref{criticalad1}. 
Since this is the difference of tortoise coordinates for the difference between $R$ and $\tilde{R}$, the derivative $r^{\ast}$ with respect to $R$ is 
\begin{align}
\frac{dr^{\ast}(R)}{dR}=\frac{\ell^2}{2 R^2}\left(\frac{2 r R}{R^2-r^2}+\log \left(\frac{R+r}{R-r}\right)\right)>0 \quad (\mbox{for $r < R$})
\end{align}
Thus, we see that $r^{\ast}$ increases by increasing $R$ at $r=r_f < R$, and therefore $\tilde{r}^{\ast}(r_s) > r^{\ast}(r_s)$ and thus, $\Delta \tau_{\infty}$ is negative.

\begin{figure}[h]
\centering
\includegraphics[keepaspectratio,scale=0.50]{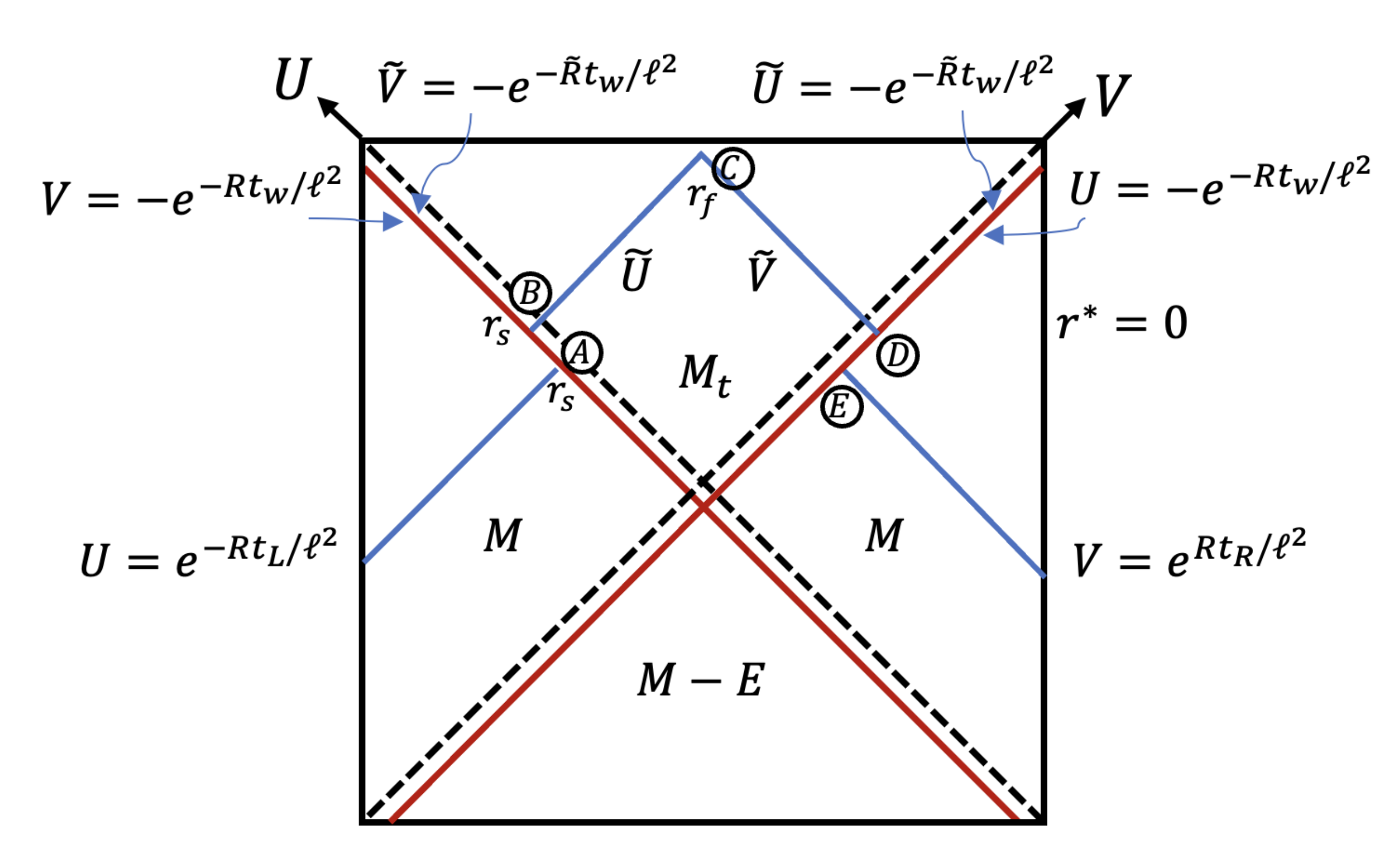}
\caption{Future boundary of WdW patch in the case of a double shock wave in AdS. Double shock waves (red line) divide the region into left, right, above, and below. We use $(U, V)$ coordinates for left and right and $(\tilde{U}, \tilde{V})$ on the above region. $M_t$ in $(\tilde{U}, \tilde{V})$ coordinates is determined by the DTR condition.}
\label{fig:AdS-Comp2}
\end{figure}

We can also examine $\tau_{\infty} < 0$ in the double scaling limit \eqref{BTZDS}. 
For that purpose, we set $\tilde{R}=\sqrt{1+\frac{E}{M}}R$. 
Furthermore, we use 
\begin{align}
r_s=R\tanh \frac{R}{2 \ell^2}(t_w+t_L)
\end{align}
This can be obtained by solving \eqref{UVcondsing2} for $r_s$. 
In the double scaling limit \eqref{BTZDS}, 
\begin{align}
R^2 - r_s^2 &\to 4 R^2 e^{- \frac{R}{\ell^2} (t_w+t_L)} \,, \\
\frac{E}{M} \frac{R^2}{R^2 - r_s^2} &\to \frac{E}{4 M} e^{+ \frac{R}{\ell^2} (t_w+t_L)} = e^{+ \frac{R}{\ell^2} t_L} \alpha \,.
\end{align}
Thus, we obtain the following
\begin{align}
\Delta \tau_{\infty} = r^{\ast}(r_s)-\tilde{r}^{\ast}(r_s)  \to 
-\frac{\ell^2}{2 R}\alpha + O(\alpha^2)<0 \,.
\end{align}
Thus $\Delta \tau_{\infty}$ is negative. \\
{\bf Double shock wave}\\
Next, we calculate the critical time in the case where two shock waves are inserted from both boundaries. For simplicity, we insert two shock waves in such a way that the resultant Penrose diagram is symmetric between left and right, see figure \ref{fig:AdS-Comp2}. Note that, as we did in section \ref{sec:2.2.1} and figure \ref{fig:AdSdshock}, there are two ways to put shock wave in the symmetry, but this time we consider the situation as shown in figure \ref{fig:AdS-Comp2} in order to perform the calculation unambiguously. 
Then we can use $(U, V)$ for the left and right region, which is separated by double shock waves (red line in the figure).

Again,  
$U=$ constant and $V=$ constant null lines (blue line) are determined by the left boundary time $t_L$ and right boundary time $t_R$. As in the single shock wave case, there are two different coordinate descriptions for a single shock wave, such as \eqref{singleshockBTZ}. However, $\tilde{R}=8 G_N M_t\ell^2$ now. As in the single shock wave case, the following conditions are derived using eq.~\eqref{UVandrcond} for points A and E, and \eqref{UVandrcond2} for points B, C, D,  
\begin{align}
\label{UVcondA2}
-  e^{-\frac{R t_L}{\ell^2}} e^{-\frac{Rt_w}{\ell^2}} &= - e^{\frac{ 2R r^{\ast}(r_s)}{\ell^2}} \quad \mbox{(point A)}\\
\label{UVcondB2}
- \tilde{U}e^{- \frac{ \tilde{R}t_w}{\ell^2}} &=- e^{\frac{ 2\tilde{R} \tilde{r}^{\ast}(r_s)}{\ell^2}} \quad \mbox{(point B)}\\
\label{UVcondC2} 
\tilde{U} \tilde{V} &=e^{\frac{2\tilde{R} \tilde{r}^{\ast}(r_f)}{\ell^2}} \hspace{-0.3mm}\quad \,\,\, \, \mbox{(point C)} \\
\label{UVcondD2} 
- e^{- \frac{\tilde{R} t_w}{\ell^2} } \tilde{V}&=- e^{\frac{2\tilde{R} \tilde{r}^{\ast}(r_s)}{\ell^2}} \hspace{-0.3mm}\quad \mbox{(point D)} \\
\label{UVcondE2} 
- e^{- \frac{R t_w}{\ell^2} }e^{\frac{R t_R}{\ell^2}}&=- e^{\frac{2 R {r}^{\ast}(r_s)}{\ell^2}} \hspace{-0.3mm}\quad \mbox{(point E)} 
\end{align}
where $\tilde{U}$ is the coordinate value on the null line between point B and C, and $\tilde{V}$ is the one between C and D. 
Just as a single shock wave case, in the double scaling limit \eqref{BTZDS}, the effect of shock waves is simply the coordinate shift in $U$ and $V$ by $\alpha$. Again setting $-t_L = t_R \equiv \tau$, and from the symmetry, $\tilde{U} = \tilde{V}$, \eqref{UVcondA2} and \eqref{UVcondE2}, and \eqref{UVcondB2} and \eqref{UVcondD2} are the same. Therefore 
from \eqref{UVcondA2}, 
\begin{align}
\tau - t_w &=  2r^{\ast}(r_s), 
\end{align}
and from \eqref{UVcondB2} and \eqref{UVcondC2}, 
\begin{align}
 \tilde{U} = e^{\frac{2 \tilde{R} \tilde{r}^{\ast}(r_s) + \tilde{R} t_w}{\ell^2}} = e^{\frac{ \tilde{R} \tilde{r}^{\ast}(r_f)}{\ell^2}}
\quad &\Rightarrow \quad t_w+2\tilde{r}^{\ast}(r_s)=\tilde{r}^{\ast}(r_f)\,.
\end{align}
From these, we obtain
\begin{align}
\tau =  2r^{\ast}(r_s) -  2\tilde{r}^{\ast}(r_s) +\tilde{r}^{\ast}(r_f)
\end{align}
Since $r_f \to 0$, $ \tilde{r}^{\ast}(r_f) \to 0$,
we obtain 
\begin{align}
\Delta \tau_{\infty}=2\left(r^{\ast}(r_s)-\tilde{r}^{\ast}(r_s)\right) \,.
\end{align}
This is twice as large as \eqref{criticalad1}. 
Thus in the double scaling limit \eqref{BTZDS}, 
\begin{align}
\label{tauAdSpert}
\Delta \tau_{\infty} = - \frac{\ell^2}{R} \alpha + O(\alpha^2) < 0\,.
\end{align}

Let us understand this in terms of constant shift. By using formula \eqref{UVcondC2}, we have at point C,
\begin{align}
\tilde{U}\tilde{V}= 1  
\end{align}
where we take $r_f \to 0$. 
Using eq.~\eqref{UtildeUrela}\footnote{Note that $\tilde{U}$ and $U$ are switched between the right figure \ref{fig:AdS-coordinate} and \ref{fig:AdS-Comp2}.}, $\tilde{U}=   U  + \alpha$, and left right symmetry of the Penrose diagram, $\tilde{V}=\tilde{U}$. 
After setting $-t_L=t_R=\tau$, we can solve this for $\tau$, and we obtain 
\begin{align}
U = e^{- \frac{R t_L}{\ell^2}}= 1 - \alpha \quad \Rightarrow \quad \tau=\frac{\ell^2}{R}\log(1-\alpha) = -\frac{\ell^2}{R}\alpha + O(\alpha^2)<0
\end{align}
This is the same quantity as in \eqref{tauAdSpert}. Here, in the last equality, we used the fact that the solution of the shift is satisfied in the limit where $\alpha$ is small.
%
\subsection{de Sitter case}
Now let us discuss complexity in the dS case. As we said earlier, we consider the stretched horizon as a holographic screen \cite{Susskind:2021esx}. Furthermore, we consider the effects of the shock waves which are inserted at times $t_L=- t_w$ on the north pole (left patch $r=0$). In the case of dS, even without the shock wave, the critical time for complexity divergence is nontrivial.

As BTZ case, in dS, when $U=$ constant and $V=$ constant null lines intersect at a certain radius $r$, the following holds 
\begin{align}
\label{UVandrconddS}
UV &= \pm e^{ - \frac{2 L r^{\ast}(r)}{\ell^2}} \,, \\
\tilde{U} \tilde{V} &= \pm e^{- \frac{2\tilde{L} \tilde{r}^{\ast}(r)}{\ell^2}}
\label{UVandrconddS2}
\end{align}
For $\pm$, we choose $-$ if $r < L$, and $+$ if $r > L$. 
$ \tilde{r}^{\ast}(r)$ is a tortoise coordinate as eq.~\eqref{tortoisedS} but in terms of $\tilde{L}$ instead of $L$. Eq.~\eqref{UVandrconddS} is valid in $(U, V)$ coordinate patch, and \eqref{UVandrconddS2} is in $(\tilde{U}, \tilde{V})$ coordinate patch.
\\\
\vspace{2mm}
\\
{\bf No shock wave}\\
First, let us calculate the critical time without a shock wave. The discussion in this case is based on \cite{Jorstad:2022mls}.
\begin{figure}[h]
\hspace{-5mm}
\centering
\includegraphics[keepaspectratio,scale=0.45]{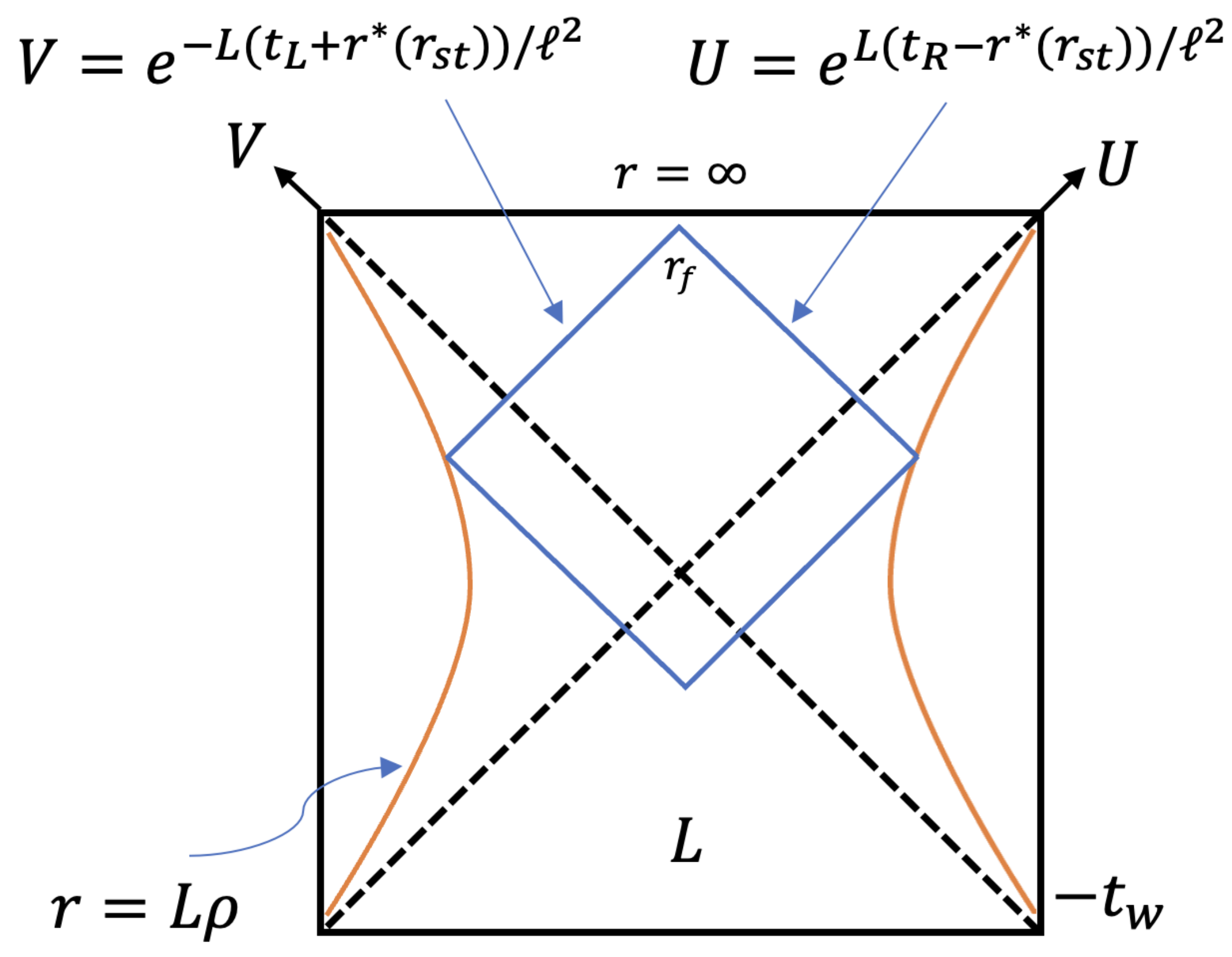}
\caption{The WdW patch is anchored by the time on the stretched horizon, where $t_L$ and $t_R$ is the time on the stretched horizon $r = \rho \ell$, where $\rho \to 1$. If this WdW patch reaches the depth of $r_f$, the critical time is the stretched horizon time when $r_f \to \infty$.}
\label{fig:dS-Comp1}
\end{figure}
\\
In figure \ref{fig:dS-Comp1}, the point $r_f$ is the location where $U$ and $V$ are determined on the stretched horizon.  
At the $r=r_f > L$ point, using eq.~\eqref{UVandrconddS} we have 
\begin{align}
U V  
 = e^{- \frac{ L \left( t_L - t_R + 2 r^{\ast} (r_{st} \right) }{\ell^2}} =e^{- \frac{2 L r^{\ast}(r_f)}{\ell^2}} 
\end{align}
Note, however, in dS, the coefficient in front of $r^{\ast}$ on the right-hand side is negative. Therefore, it follows that
\begin{align}
{-t_L+t_R-2r^{\ast}(r_{st})}=-2 r^{\ast}(r_f)\,.
\end{align}
Setting $-t_L=t_R=\frac{\ell^2}{L} \tau$ and considering the pure dS case, $L = \ell$, and we obtain the critical time $\tau = \tau_\infty^{(0)}$ when $r_f \to \infty$ as, 
\begin{align}
\ell \tau_\infty^{(0)} &= r^{\ast}(r_{st})-r^{\ast}(r_f \to \infty)  \to r^{\ast}(r_{st})\,, \nonumber \\
\tau_{\infty}^{(0)} &= \frac{ r^{\ast}(r_{st}) }{\ell} = \frac{1}{2} \log \left|\frac{1+\rho}{1-\rho} \right|={\rm Arctanh} \rho \,.
\label{noshocktau0}
\end{align}
This is the same result as in \cite{Jorstad:2022mls}. Here $\tau^{(0)}_\infty$ represents the critical time without shock waves. \\\ \\
{\bf Single shock wave}\\
Next, let us consider the case where a shock wave is inserted at the time $t_L=-t_w$ on the north pole.  As shown in Section \ref{sec:shockgeodS}, in this case, the lower part of the Penrose diagram is pure dS.
\begin{figure}[h]
\hspace{-17mm}
\includegraphics[keepaspectratio,scale=0.48]{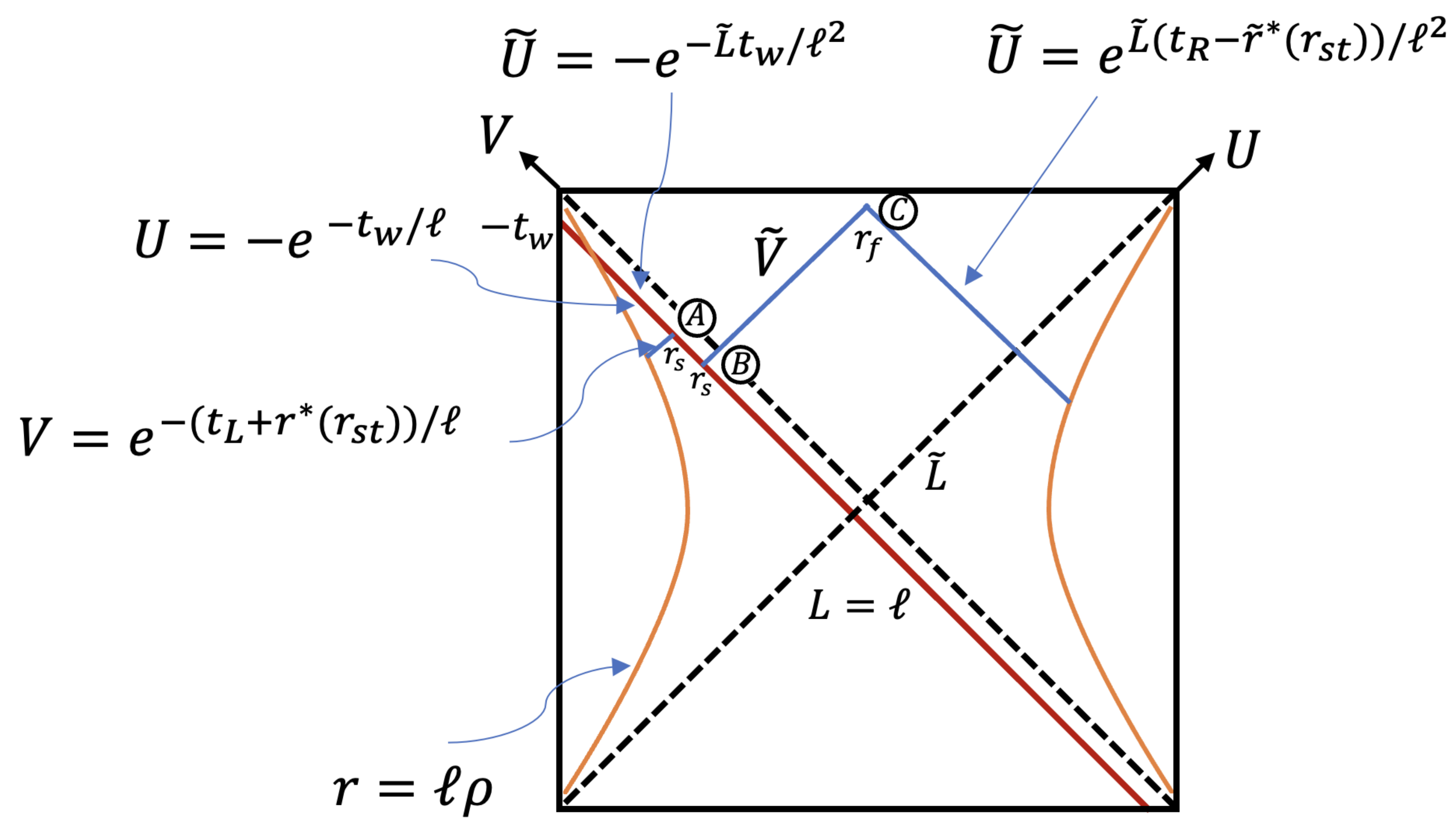}
\caption{Consider the future boundary of the WdW patch in dS. If this patch reaches the depth of $r_f$ (= point C), the critical time is the boundary time when $r_f=\infty$. The stretched horizon is $r_{st} = \rho \, \ell$, with $\rho \to 1$. }
\label{fig:dS-Comp2}
\end{figure}
\ \\
Similarly in this case, the following equation holds  eq.~\eqref{UVandrconddS} for point A and \eqref{UVandrconddS2} for point B and C, 
\begin{align}
\label{dSUVcondA}
-  e^{-\frac{ t_w}{\ell}} V &= - e^{- \frac{ 2  {r}^{\ast}(r_s)}{\ell}} \quad \,\, \mbox{(point A)}\\
\label{dSUVcondB}
- e^{- \frac{ \tilde{L} t_w}{\ell^2}} \tilde{V} &= - e^{- \frac{ 2\tilde{L} \tilde{r}^{\ast}(r_s)}{\ell^2}} \quad \mbox{(point B)}\\
\label{dSUVcondC} 
e^{\frac{  \tilde{L} \left(t_R - r^{\ast}(r_{st}) \right) }{\ell^2}} \tilde{V} &=e^{ - \frac{2\tilde{L} \tilde{r}^{\ast}(r_f)}{\ell^2}} \hspace{-0.3mm}\quad \,\,\,\, \mbox{(point C)}
\end{align}
where $\tilde{V}$ is the coordinate value on the null line between points B and C, and we have 
\begin{align}
\label{valueofVdS}
V = e^{- \frac{ t_L + r^{\ast}(r_{st})}{\ell}}
\end{align}
for the coordinate value on the null line between point A.

Using eq.~\eqref{dSUVcondA} and \eqref{dSUVcondB}, 
in this case too, the shift can be understood under the limit as follows.
\begin{align}
\tilde{V}-V &= e^{\frac{\tilde{L} \left(t_w - 2 \tilde{r}^{\ast}(r_s) \right) }{\ell^2} }   - e^{\frac{t_w - 2 r^{\ast}(r_s)}{\ell} }    \\
& = e^{\tilde{L} t_w/\ell^2} \frac{\tilde{L}-r_s}{\tilde{L}+r_s}  -e^{t_w/\ell} \frac{\ell-r_s}{\ell+r_s}
\end{align}
Substituting $\tilde{L}= \ell \sqrt{1-8G_N E}$,  and taking the double scaling limit \eqref{DSdS}, with $r_s \to \ell$, we obtain 
\begin{align}
\label{dsBeta2VtV}
\tilde{V}-V \to   -2G_N Ee^{t_w/\ell}  = - \beta
\end{align}
which matches the \eqref{dSbeta}. 

The calculation of the critical time is straightforward, however, it must be clear from the figure \ref{fig:dS-Comp1} that the critical time is {\it greater}  than the one without a shock wave. Using eq.~\eqref{dSUVcondA} with \eqref{valueofVdS}, and the eq.~\eqref{dSUVcondB} divided by \eqref{dSUVcondC} become, respectively 
\begin{align}
\label{dSseveraltrela1}
& t_L+t_w   + r^{\ast}(r_{st})  =2r^{\ast}(r_s)  \\
\label{dSseveraltrela2}
&  t_R + t_w - r^{\ast}(r_{st}) = - 2 \tilde{r}^{\ast}(r_f) + 2 \tilde{r}^{\ast}(r_s) 
\end{align}
Eliminating $t_w$ from the above and setting $-t_L=t_R=\ell \tau$, and using at $r_f \to \infty$, $\tilde{r}^{\ast}(r_f) \to 0$, we obtain the following 
\begin{align}
\label{Delay1}
\ell \tau_{\infty}=r^{\ast}(r_{st})+\tilde{r}^{\ast}(r_s)-r^{\ast}(r_s)
\end{align}
The term $r^{\ast}(r_{st})$ is the critical time without shock waves, \eqref{noshocktau0}, therefore  
$\ell \tau_{\infty}^{(0)} =r^{\ast}(r_{st})$, and  
the delay of critical time is given by 
\begin{align}
\label{DeltataudiffdS}
  \Delta \tau_{\infty} \equiv \tau_{\infty}- \tau_{\infty}^{(0)}
  = \frac{1}{\ell}\left( \tilde{r}^{\ast}(r_s)-r^{\ast}(r_s)\right)
\end{align}

To understand what eq.~\eqref{DeltataudiffdS} implies, we need to consider the difference of tortoise coordinate between $L$ and $\tilde{L}$. In our setting $r_s < L = \ell$, and $\tilde{L} = \ell \sqrt{1 - 8 G_N E}$. 
Since 
\begin{align}
\frac{d \tilde{r}^{\ast}(r_s)}{dE}  &= \frac{d \tilde{L}}{dE} \times \frac{d \tilde{r}^{\ast}(r_s)}{d\tilde{L}} \nonumber \\
& = - \left( \frac{4 \ell G_N}{\sqrt{1 -8 G_N E}}\right)  \times  - \left( \frac{ \ell^2 r_s}{L (L^2 - r_s^2)} + \frac{\ell^2}{2 L^2} \log \frac{L+r_s}{L-r_s}\right) > 0
\label{dSderivativeargument}
\end{align}
Therefore, the critical time becomes greater by the shock wave.

\begin{figure}[h]
\hspace{-8mm}
\includegraphics[keepaspectratio,scale=0.45]{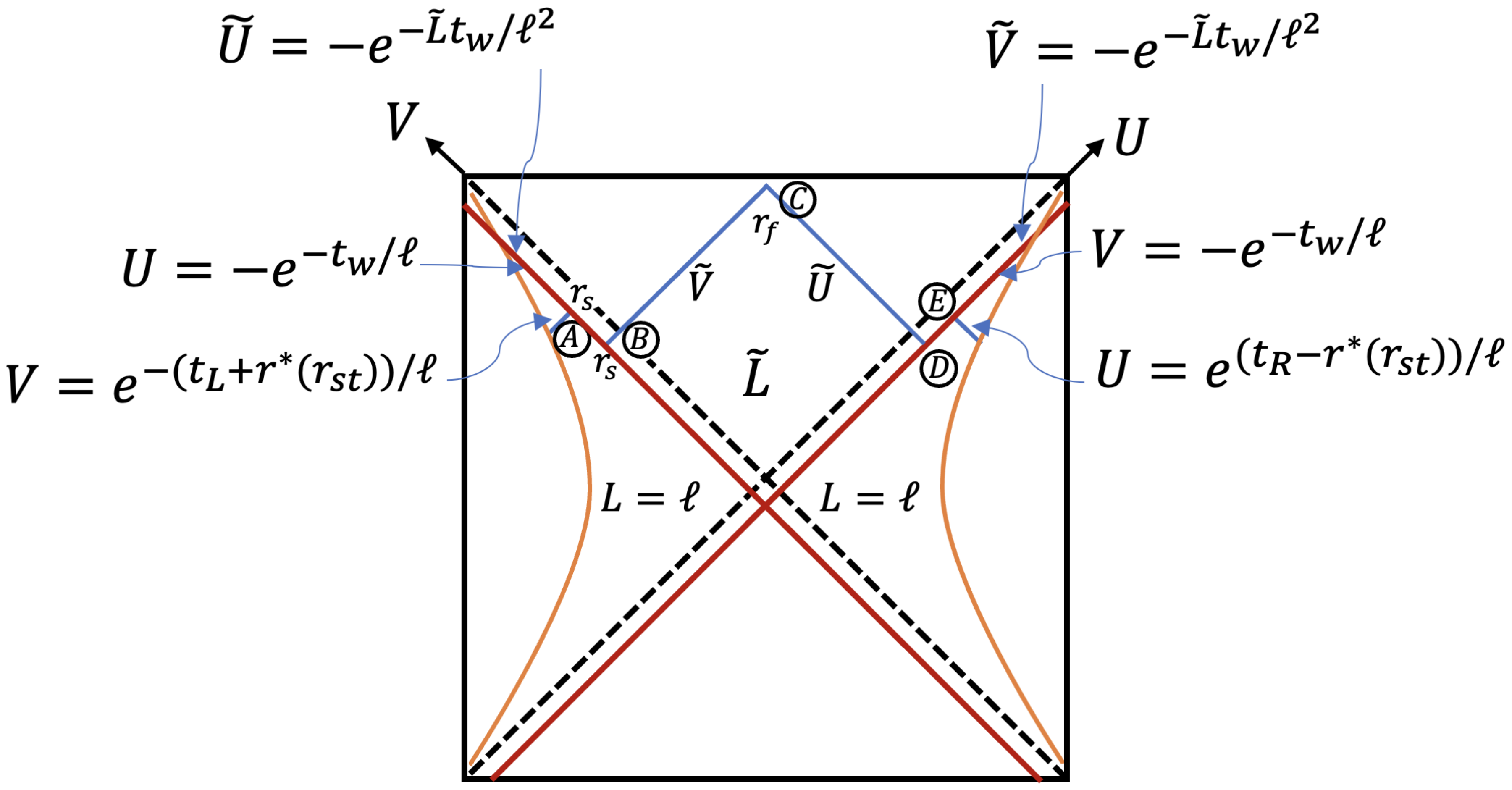}
\caption{Future boundary of WdW patch in the case of a double shock wave in dS. The black hole mass of the below part is determined by the DTR condition.}
\label{fig:dS-Comp3}
\end{figure}

In the double scaling limit \eqref{DSdS}, $r_s \to L$. To analyze this carefully,  using \eqref{dSseveraltrela1}, 
\begin{align}
& t_L+t_w   + r^{\ast}(r_{st})  =2r^{\ast}(r_s) =  \ell \log \frac{\ell + r_s}{\ell - r_s} \,\\
& \Rightarrow \quad r_s ={\ell} \, {\tanh \frac{t_L+t_w+ r^{\ast}(r_{st})}{2 \ell}} \,.
\end{align}
Therefore in the double scaling limit \eqref{DSdS}, we can derive the delay of critical time as 
\begin{align}
\label{taudSpert}
\Delta \tau_{\infty}  
\to \frac{\beta}{2} + O(\beta^2) >0 \,.
\end{align}
Thus $\Delta \tau_{\infty}$ is positive, and this implies the delay of hyperfast growth for small $\beta$ perturbation. 
\\\ \\
{\bf Double shock waves}\\
Finally, we compute the case of a dS with a double shock wave. As shown in Figure \ref{fig:dS-Comp3}, we insert double shock waves and consider the critical time at which the future boundary of the WdW patch extended from a certain time on the stretched horizon reach infinity, {\it i.e.,} $r_f=\infty$ of the point C.

As usual, the following is satisfied, 
\begin{align}
\label{dSUVcondA2}
-  e^{-\frac{ t_w}{\ell}}   e^{- \frac{ t_L + r^{\ast}(r_{st})}{\ell}}&= - e^{- \frac{ 2  {r}^{\ast}(r_s)}{\ell}} \quad \,\, \mbox{(point A)}\\
\label{dSUVcondB2}
- e^{- \frac{ \tilde{L} t_w}{\ell^2}} \tilde{V} &= - e^{- \frac{ 2\tilde{L} \tilde{r}^{\ast}(r_s)}{\ell^2}} \quad \mbox{(point B)}\\
\label{dSUVcondC2} 
\tilde{U} \tilde{V} &=e^{ - \frac{2\tilde{L} \tilde{r}^{\ast}(r_f)}{\ell^2}} \hspace{-0.3mm}\quad \,\,\,\, \mbox{(point C)} \\
\label{dSUVcondD2}
- e^{- \frac{ \tilde{L} t_w}{\ell^2}} \tilde{U} &= - e^{- \frac{ 2\tilde{L} \tilde{r}^{\ast}(r_s)}{\ell^2}} \quad \mbox{(point D)}\\
\label{dSUVcondE2}
-  e^{-\frac{ t_w}{\ell}} e^{ \frac{ t_R - r^{\ast}(r_{st})}{\ell}} &= - e^{- \frac{ 2  {r}^{\ast}(r_s)}{\ell}} \quad \,\, \mbox{(point E)}
\end{align}
where $\tilde{V}$ is the coordinate value on the null line between point B and C, and $\tilde{U}$ is the one between C and D. 
Just as a single shock wave case, in the double scaling limit \eqref{DSdS}, the effect of shock waves is simply the coordinate shift in $U$ and $V$ by $\beta$. Again setting $-t_L = t_R $, and from the left-right symmetry of the Penrose diagram, $\tilde{U} = \tilde{V}$, \eqref{dSUVcondA2} and \eqref{dSUVcondE2}, and \eqref{dSUVcondB2} and \eqref{dSUVcondD2} are the same. From eq.~\eqref{dSUVcondA2} - \eqref{dSUVcondC2}, we obtain 
\begin{align}
t_L+t_w  &=2r^{\ast}(r_s)- r^{\ast}(r_{st})\\
  t_w &=2\tilde{r}^{\ast}(r_s)  - \tilde{r}^{\ast}(r_f)
\end{align}
Setting $-t_L = t_R \equiv \ell {\tau}$ and using $\tilde{r}^{\ast}(r_f) \to 0$ as $r_f \to \infty$, 
\begin{align}
\label{finalresult1}
\Delta \tau_{\infty} = \tau_{\infty} - \frac{r^{\ast}(r_{st})}{\ell}&=\frac{2}{\ell}\left(\tilde{r}^{\ast}(r_s)-r^{\ast}(r_s)\right) \\
\label{finalresult1-2}
&\to \beta + O(\beta^2)  > 0 \,.
\end{align}
Thus, as in the case of AdS, the change in critical time is twice as large as a single shock wave case.  
Here, we used the fact that the double shock wave solution is satisfied in the limit where $\beta$ is small. 

Let us understand this result in terms of constant shift of the metric in eq.~\eqref{shockgeo2} as BTZ case.  By using formula \eqref{dSUVcondC2}, at $r=r_f \to \infty$ we have 
\begin{align}
\tilde{U} \tilde{V}\to  1 \,, 
\end{align}
From \eqref{dSbeta} (or equivalently \eqref{dsBeta2VtV}), 
\begin{align}
\tilde{V} = V - \beta
\end{align}
Using the left-right symmetry of the Penrose diagram, we obtain $\tilde{U}=\tilde{V}$ and 
\begin{align} 
\label{VVVfinalcritical}
V  = e^{- \frac{t_L + r^\ast(r_{st})}{\ell} } \to 1  + \beta \quad \Rightarrow \quad - \frac{t_L}{\ell} \to  \frac{r^\ast(r_{st})}{\ell} + \log (1 +\beta) \,.
\end{align}
Setting $-t_L=t_R=\ell \tau$, we obtain the critical time $\tau_{\infty}$ as the right hand side of \eqref{VVVfinalcritical} and therefore, 
\begin{align}
\label{finalresult2}
\Delta \tau_\infty  =  \tau_\infty - \frac{ r^\ast(r_{st}) }{\ell} = \log(1+ {\beta}) = {\beta} + O(\beta^2) >0
\end{align}
This is the same quantity as in \eqref{finalresult1}. 

%
\subsection{CV calculation}
Finally, let us derive the above critical time delay based on CV calculation as well. 
For that purpose, we need to calculate the volume of the extremal geodesic anchored to a time slice on the stretched horizon. The following discussion is based on geodesic surface prescriptions between different patches, as is worked out in \cite{Chapman:2018dem,Chapman:2018lsv}. We are interested in three dimensions. However, we will proceed a little in the general dimension and finally discuss the three-dimensional case.

In the Eddington-Finkelstein coordinate, the metric for the left (right) static patch and the future patch is written in terms of $v_L$ ($u_R$) as 
\begin{align}
ds^2=-f(r)dv_L^2 +2dv_Ldr+r^2 d\Omega^2_{d-1} 
\left( =-f(r)du_R^2 -2du_Rdr+r^2 d\Omega^2_{d-1} \right) \,.
\end{align}
Then the volume of the geodesic is given by
\begin{align}
\label{CVvolume}
V=\Omega_{d-1} \int dsr^{d-1}\sqrt{-f\dot{v}_L^2+2\dot{v}_L \dot{r}}
\left( =\Omega_{d-1} \int dsr^{d-1}\sqrt{-f\dot{u}_R^2-2\dot{u}_R \dot{r}} \right)
\end{align}
where the dot is derivative with respect to geodesic parameter $s$. $\Omega_{d-1}$ is the volume of a unit $(d-1)$-dimensional sphere. To simplify the notation, we sometimes omit $L$ and $R$ in $v_L$ and $u_R$.

The shock wave separates spacetime into several patches, as a result, different $f(r)$ is assigned for each spacetime patch. However, there is a conserved quantity in each patch. If we regard \eqref{CVvolume} as the action, its Lagrangian is independent of $v$ (or $u$). Therefore the following $P$ is conserved with respect to $s$; 
\begin{align}
& P \equiv \frac{1}{\Omega_{d-1}} \frac{\delta V}{\delta \dot{v}}=\frac{\left( \dot{r}-f\dot{v}  \right) r^{d-1}}{\sqrt{-f\dot{v}^2+2\dot{v} \dot{r}}}  \,, \quad \left(  P \equiv \frac{1}{\Omega_{d-1}} \frac{\delta V}{\delta \dot{u}}=\frac{\left( -\dot{r}-f\dot{u}  \right) r^{d-1}}{\sqrt{-f\dot{u}^2- 2\dot{u} \dot{r}}} \right) \,.
\end{align}
Since \eqref{CVvolume} is invariant under the reparametrization for parameter $s$, 
we can always choose a parameterization gauge such that 
\begin{align}
\label{CVeq1}
\sqrt{-f\dot{v}^2+2\dot{v} \dot{r}}=r^{d-1} \,, \quad \left( \sqrt{-f\dot{u}^2-2\dot{u} \dot{r}}=r^{d-1}  \right) \,. 
\end{align}
Then we have 
\begin{align}
\label{CVeq2}
P =  \dot{r}-f\dot{v}  \,, \quad \left(P =  -\dot{r}-f\dot{u}  \right) \,.
\end{align}

Consider a geodesic line extending from the left stretched horizon to the right stretched horizon. $r$ extends from the left stretched horizon to a point called the turning point $r= r_{turn}$. After that, $r$ decreases to the right boundary. $r$ is a continuous parameter on the shock wave due to the $C^0$ nature of the metric and we set $r = r_s$ at the point where the geodesic line crosses the shock wave as before.  

\begin{figure}[t]
\centering
\includegraphics[keepaspectratio,scale=0.41]{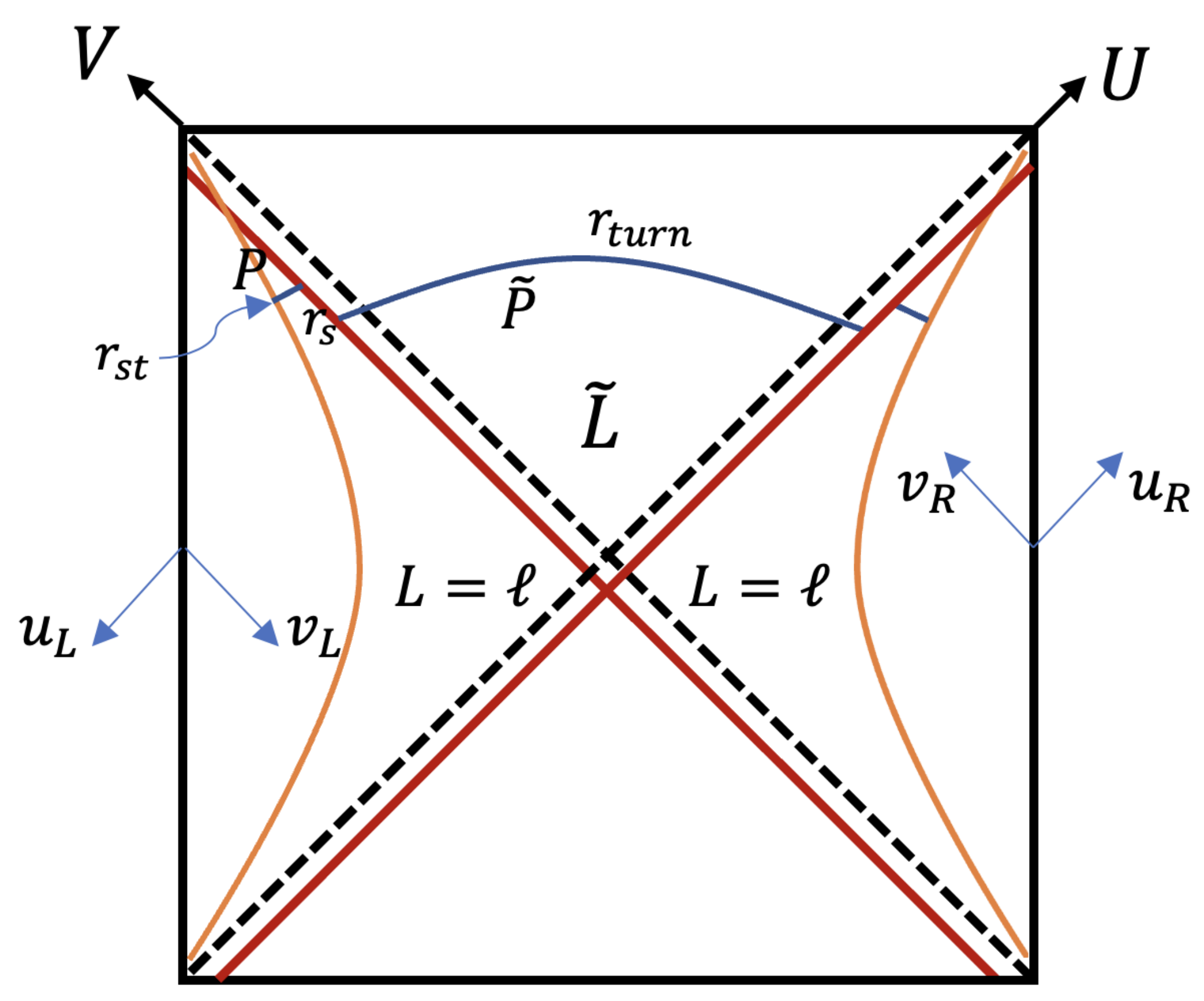}
\caption{Figure of a geodesic line. The turning point is assumed to be in the interior of pure dS. The momentum of pure dS is assumed to be $P$, and that of the SdS patch is assumed to be $\tilde{P}$. 
}
\label{fig:dS-CVarg}
\end{figure}

As we have mentioned, $P$ is conserved only within each patch. Therefore we denote $P$ in the patch where the cosmological horizon is $L$, and $\tilde{P}$ for $\tilde{L}$. Similarly we put $\tilde{\,}$, 
for the quantity in $\tilde{L}$, such as $\tilde{f}(r)$, $\tilde{u}$, $\tilde{v}$, etc.    
See figure \ref{fig:dS-CVarg}. 

Eq.~\eqref{CVeq1} and \eqref{CVeq2} give 
\begin{align}
\label{CVeq3}
\dot{r} &=  + {\sqrt{P^2  + r^{2d-2} f(r) }} \,,  \qquad \left(   \dot{r} =  - {\sqrt{P^2  + r^{2d-2} f(r) }} \right) \,, \\
\dot{v} & = \frac{1}{f(r)} \left( -P + \sqrt{ P^2 + r^{2d -2} \, f(r)   } \right)  \,, \quad  \Bigl(  \dot{u} = \frac{1}{f(r)} \left( -P + \sqrt{ P^2 + r^{2d -2} \, f(r)   } \right) \Bigr) \,,
\label{CVeq4}
\end{align}
where we choose the convention that $s$ increase starting from left stretched horizon toward $r_{turn}$. 
With this, the volume is given by 
\begin{align}
\mathcal{C}_V &= \Omega_{d-1} \int ds r^{2(d-1)} \nonumber \\
&= 2 \Omega_{d-1} \left( \int_{r_{st}}^{r_s} \frac{r^{2(d-1)}}{\sqrt{P^2+f(r)r^{2(d-1)}}}dr+\int_{r_s}^{r_{turn}} \frac{r^{2(d-1)}}{\sqrt{\tilde{P}^2+\tilde{f}(r)r^{2(d-1)}}}dr \right) \,.
\end{align}

Turing point is defined as the point where the $\dot{r}$ is zero,
\begin{align}
\label{rturntoPdef}
\left(1-\frac{r_{turn}^2}{\tilde{L}^2} \right)r_{turn}^{2(d-1)} + \tilde{P}^2=0
\end{align}
As is clearly seen from this equation, the turning point $r_{turn} \to \infty$ as $ \tilde{P} \to \infty$. Therefore, 
we will investigate $ \tilde{P} \to \infty$ limit.  In fact, in this limit, $P \to \infty$ as well because the effect of the shock wave in the double scaling limit \eqref{DSdS} is simply a coordinate shift\footnote{Under the coordinate shift, if momentum diverges in one patch, it also diverges in another patch as well. More formally, one can obtain the relation between $ \tilde{P}$ and $P$ as follows. 1) Deriving the equation of motion for $r(s)$ from \eqref{CVvolume}  
as $\ddot{r} =\frac{\dot{v}^2}{2}\partial_v f+\frac{1}{2}\partial_r\left( r^{2d-2}f \right)$. 
2) By integrating out this equation with respect to $s$ between $r = r_s - \epsilon$ to $ r_s + \epsilon$  with  $\epsilon \to 0$, we obtain how $\dot{r}$ jumps by crossing the shock wave. 3) By furthermore using  this with \eqref{CVeq2}, \eqref{CVeq3} and \eqref{CVeq4}, we obtain the relation between $ \tilde{P}$ and $P$. By taking $ \tilde{P} \to \infty$, one can see in fact that $P \to \infty$ as well.   
}.

Using \eqref{CVeq3} and \eqref{CVeq4}, the variation of $u$ and $v$ along the geodesic can be calculated by the following integral
\begin{align}
\Delta v =\int \frac{\dot{v}}{\dot{r}} dr=  \int T( P,r)dr \,, \quad 
\left( \Delta u =\int \frac{\dot{u}}{\dot{r}} dr =-\int T( P,r)dr \right) \,,
\end{align}
where $T(P,r)$ is 
\begin{align}
\label{definitionofTPr}
T(P,r)=\frac{1}{f(r)}-\frac{P}{f(r)\sqrt{P^2 + f(r)r^{2(d-1)}}} \,.
\end{align}
Consider a geodesic that leaves the left-stretched horizon and reaches the right-stretched horizon. 
From the left stretched horizon to $r_{turn}$, we use $v_L$ and $\tilde{v}_L$, and from $r_{turn}$ to the right stretched horizon, we use $u_R$ and $u_R$, as 
\begin{align}
\label{CVeqv1}
v_L(r_s) \,-v_L(r_{st}) &=  + \int_{r_{st}}^{r_s} T(P,r)dr \,, \\
\label{CVeqv2}
\tilde{v}_L(r_{turn})- \tilde{v}_L(r_s)  &= + \int_{r_s}^{r_{turn}}T(\tilde{P},r)dr \,, \\
\label{CVequ2}
\tilde{u}_R(r_{turn}) - \tilde{u}_R(r_s) &=-\int_{r_{s}}^{r_{turn}}T(\tilde{P},r)dr \,, \\
\label{CVequ1}
u_R(r_s)\, - u_R(r_{st})  &=-\int_{r_{st}}^{r_{s}} T(P,r)dr \,,
\end{align}
Note that $v_L(r_s) \neq \tilde{v}_L(r_s)$ since $v_L$ jump at the shock wave\footnote{In \eqref{Vaidya2dSLLt}, we use $u_L$ which covers left and past patch, where  $u_L$ does not jump but $v_L$ jumps at the shock wave.}. 
In fact, from the definitions of $u$ and $v$ in eq.~\eqref{dSuvcoord}, we have 
\begin{align}
\Delta v (r_s) \equiv  \tilde{v}_L(r_s) - v_L(r_s)  &=  \left( \tilde{u}_L(r_s) + 2 \tilde{r} ^\ast(r_s) \right)  - \left( u_L(r_s) + 2 r^\ast(r_s) \right) \nonumber \\
&= 2 \left(    \tilde{r}^\ast(r_s)  - r^\ast(r_s) \right)  \,.
\end{align}
Here we use the continuity of $u_L$ on the $u$ = constant shock wave (= shock wave propagating $V$ direction). 
Similarly $u_R(r_s) \neq \tilde{u}_R(r_s)$ and 
\begin{align}
 \tilde{u}_R(r_s) - u_R(r_s)   &= \left( \tilde{v}_R(r_s)  - 2 \tilde{r} ^\ast(r_s)  \right) -  \left( v_R(r_s) -2 r^\ast(r_s) \right)  \nonumber \\
& = -2 \left( \tilde{r}^\ast(r_s)  - r^\ast(r_s)   \right)  =   - \Delta v (r_s) \,.
\end{align} 
due to the continuity of $v_R$ on the $v$ = constant shock wave  (= shock wave propagating $U$ direction). 

Adding \eqref{CVeqv1} with  \eqref{CVeqv2}, and \eqref{CVequ1} with \eqref{CVequ2}, we have 
\begin{align}
\label{CVeqv3}
\tilde{v}_L(r_{turn})-v_L(r_{st}) - \Delta v(r_s) &=\int_{r_{st}}^{r_s} T(P,r)dr+\int_{r_s}^{r_{turn}}T(\tilde{P},r)dr \,, \\
\tilde{u}_R(r_{turn}) - u_R(r_{st}) + \Delta v(r_s) &= - \int_{r_{st}}^{r_s} T(P,r)dr- \int_{r_s}^{r_{turn}}T(\tilde{P},r)dr \,, 
\label{CVequ4}
\end{align}
Using the definitions of $u$ and $v$, we have
\begin{align}
u_R(r_{st})- \tilde{u}_R(r_{turn}) + \tilde{v}_L(r_{turn})-v_L(r_{st})  
= t_R  - t_L - 2 r^\ast(r_{st})+ 2 \tilde{r}^{\ast}(r_{turn}) 
\end{align}
Thus, by subtracting \eqref{CVequ4} from \eqref{CVeqv3}, we have 
\begin{align}
\label{semifinalrelation}
 t_R  - t_L - 2 r^\ast(r_{st})+ 2 \tilde{r}^{\ast}(r_{turn}) - 2 \Delta v(r_s) 
 = 2 \int_{r_{st}}^{r_s} T(P,r)dr+2 \int_{r_s}^{r_{turn}}T(\tilde{P},r)dr \,.
\end{align}

As we have mentioned, 
in the limit where the turning point $r_{turn} \to \infty$, 
both $P$ and $\tilde{P}$ goes to infinity,  then the 
right hand side of eq.~\eqref{semifinalrelation} vanishes. 
This is because $ T(P, r) $ given \eqref{definitionofTPr} vanishes as 
\begin{align}
 T(\tilde{P}, r) = O\left( \frac{r^{2 d - 2}}{\tilde{P}^2} \right) \to 0 \quad \mbox{(at $\tilde{P} \to \infty$)} \,,
\end{align}
and from \eqref{rturntoPdef}, 
\begin{align}
\tilde{P}^2 = O\left( \left( r_{turn} \right)^{2d} \right)  \to \infty  \quad \mbox{(at $r_{turn} \to \infty$)} \,,
\end{align}
therefore 
\begin{align}
\int^{r_{turn}}  T(\tilde{P}, r)  dr = O\left( \frac{1}{r_{turn}}\right) \to 0  \quad \mbox{(at $r_{turn} \to \infty$)} \,.
\end{align}
Finally setting  $t_R = -t_L  = \ell \tau$ and using $ \tilde{r}^{\ast}(r_{turn}) \to 0$ at $r_{turn} \to \infty$, 
we have  
\begin{align}
\ell \tau_{\infty}& =  r^\ast(r_{st}) + \Delta v(r_s) =  r^\ast(r_{st})  + 2 \left( \tilde{r}^\ast(r_s) - r^\ast(r_s)   \right)
\end{align}
Thus we have 
\begin{align}
\ell \tau^{(0)}_{\infty} &=  r^{\ast}(r_{st})\,, \\
\Delta  \tau_{\infty}  &= \tau_{\infty} - \tau^{(0)}_{\infty}   =   \frac{2}{\ell} \left(  \tilde{r}^{\ast}(r_s)-r^{\ast}(r_s)  \right)   
\end{align}
which is exactly the same as eq.~\eqref{noshocktau0} and \eqref{finalresult1}. Therefore, in three dimensions, as \eqref{finalresult1-2},
\begin{align}
\Delta  \tau_{\infty} > 0 
\end{align}
In this way, even with the CV argument\footnote{In fact one can also show $\tilde{r}^{\ast}(r_s) > r^{\ast}(r_s)$ and thus $\Delta  \tau_{\infty} > 0 $ in higher dimensions as well. For example, in 3+1 dimensions, blackening factor becomes 
$f(r) = 1 - \frac{r^2}{\ell^2} - \frac{2 G_N E}{r} \sim 1 - \frac{r^2}{\ell^2} - \frac{2 G_N E}{\ell}$ since $r_s \to \ell$. Compared with the blackening factor in 2+1 dimension, this is just a formal replacement $8 G_N E \to 2 G_N E/\ell$, therefore inequality \eqref{dSderivativeargument} holds as well.}, exactly the same conclusions can be derived as WdW patch argument for CA/CV2.0.

%
\section{Discussion}
In this paper, we systematically study shock wave geometries and see how the shock waves affect the WdW patch for CA/CV2.0 complexity. We study on both BTZ and dS background, and see that the effects of shock waves are exactly opposite signs between BTZ and dS. Especially we see that the hyperfast nature of the dS complexity is always delayed, {\it i.e.}, the critical time at which complexity diverges, becomes always greater by shock wave perturbation which satisfies ANEC.

Our result is consistent with Gao and Wald's theorem \cite{Gao:2000ga} and also the analysis in \cite{Aalsma:2020aib}. If we compute the geodesic distance between the north and south pole with shock wave insertion, then unlike the BTZ case where the wormhole becomes longer, the distance becomes lesser. This reflects the point that the two-point function between the north pole and south pole grows instead of decay before the scrambling time $\ell \log S$ \cite{Aalsma:2020aib}. 
This is due to the fact that de Sitter's Penrose diagram has shrunk horizontally and the geodesic length has shortened.

As we have mentioned in the introduction, the point that the shock waves delay the hyperfast nature might not be so surprising. dS is the spacetime where the cosmological constant dominates. Once we perturb the spacetime by the shock wave insertion in the far past, then cosmological constant dominance can be destroyed due to the shock waves. However, considering this to be a characteristic of dual field theory, our results suggest something very nontrivial. This is because complexity becomes generically greater under the perturbation. 
Therefore, the results obtained here provide a restriction to the structure of the state of field theories for holographic dual to dS, which could be a double-scaled limit of SYK,  DSSYK$_\infty$ proposed by Susskind \cite{Susskind:2021esx}. 
Thus, a dual state like TFD in de Sitter must have its complexity, such that the critical time at which the complexity diverges is delayed for perturbations that correspond to shock waves in bulk spacetime.

The double-scaled limit of SYK itself is a very interesting subject by itself and recently using chord diagram technique \cite{Berkooz:2018jqr}, many interesting aspects are found recently \cite{Shaghoulian:2022fop, Lin:2022nss, Lin:2022rbf, Susskind:2022bia, Rahman:2022jsf, Susskind:2023hnj, Susskind:2023rxm}.  
To find the dS dual, the candidate of the dS dual must satisfy several properties of dS, such as the existence of the maximum entropy, type II$_1$ von Neumann algebra \cite{Chandrasekaran:2022cip}, etc. 
We wish our analysis of complexity is useful as one of these criteria and will help us understand more about dS dual in the future. 


\acknowledgments
We would like to thank Tomonori Ugajin and Nicol\'o Zenoni for their helpful discussions. We would also like to thank Sunil Sake and Nicol\'o Zenoni for their helpful comments on our draft. 
The work of TA and NI was supported in part by JSPS KAKENHI Grant Number 21J20906(TA), 18K03619(NI). The work of NI was also supported by MEXT KAKENHI Grant-in-Aid for Transformative Research Areas A “Extreme Universe” No. 21H05184.

\bibliography{paper}
\bibliographystyle{utphys}

\end{document}